\documentclass[twocolumn]{bmcart}
\usepackage{acronym}
\usepackage{amsmath}
\usepackage{amssymb}
\usepackage{color}
\usepackage{graphicx}
\usepackage[utf8]{inputenc}
\usepackage{latexsym}
\usepackage{times}
\usepackage{stmaryrd}

\newcommand{\VV}{\mathcal{V}}

\newcommand{\aap}{A\&A}
\newcommand{\apj}{ApJ}
\newcommand{\apjl}{ApJ}
\newcommand{\apjs}{ApJS}
\newcommand{\mnras}{Mon. Not. R. Astron. Soc.}
\newcommand{\pre}{Physical Review E}
\newcommand{\prl}{Phys. Rev. Lett.}

\renewcommand{\aa}{\mathbf{a}}

\newcommand{\dd}{\mathrm{d}}

\newcommand{\ii}{\mathrm{i}}
\newcommand{\kk}{\mathbf{k}}
\newcommand{\qq}{\mathbf{q}}
\newcommand{\rr}{\mathbf{r}}

\newcommand{\vv}{\mathbf{v}}
\newcommand{\xx}{\mathbf{x}}

\newcommand{\cf}{\textit{cf.,}~}
\newcommand{\ie}{\textit{i.e.,}~}
\newcommand{\eg}{\textit{e.g.,}~}

\begin{document}

\begin{frontmatter}

\begin{fmbox}
\dochead{Research}

\title{Implicit large eddy simulations of anisotropic weakly compressible
turbulence with application to core-collapse supernovae}

\author[
  addressref={caltech},
  corref={caltech},
  email={dradice@caltech.edu}
]{\inits{DR}\fnm{David} \snm{Radice}}
\author[
  addressref={caltech},
  corref={caltech}
]{\inits{SMC}\fnm{Sean M} \snm{Couch}}
\author[
  addressref={caltech},
  corref={caltech}
]{\inits{CDO}\fnm{Christian D} \snm{Ott}}

\address[id=caltech]{%
  \orgname{TAPIR, Walter Burke Institute for Theoretical Physics, California
  Institute of Technology},
  \street{E California Blvd},
  \city{Pasadena},
  \postcode{CA 91125},
  \cny{USA}
}

\begin{abstractbox}
\begin{abstract}
In the implicit large eddy simulation (ILES) paradigm, the dissipative
nature of high-resolution shock-capturing schemes is exploited to
provide an implicit model of turbulence. The ILES approach has been
applied to different contexts, with varying degrees of success. It is
the de-facto standard in many astrophysical simulations and in
particular in studies of core-collapse supernovae (CCSN). Recent 3D
simulations suggest that turbulence might play a crucial role in
core-collapse supernova explosions, however the fidelity with which
turbulence is simulated in these studies is unclear. Especially
considering that the accuracy of ILES for the regime of interest in
CCSN, weakly compressible and strongly anisotropic, has not been
systematically assessed before. Anisotropy, in particular, could
impact the dissipative properties of the flow and enhance the
turbulent pressure in the radial direction, favouring the explosion.
In this paper we assess the accuracy of ILES using numerical methods
most commonly employed in computational astrophysics by means of a
number of local simulations of driven, weakly compressible,
anisotropic turbulence. Our simulations employ several different
methods and span a wide range of resolutions. We report a detailed
analysis of the way in which the turbulent cascade is influenced by
the numerics. Our results suggest that anisotropy and compressibility
in CCSN turbulence have little effect on the turbulent kinetic energy
spectrum and a Kolmogorov $k^{-5/3}$ scaling is obtained in the
inertial range. We find that, on the one hand, the kinetic energy
dissipation rate at large scales is correctly captured even at
low resolutions, suggesting that very high ``effective
Reynolds number'' can be achieved at the largest scales of the
simulation. On the other hand, the dynamics at intermediate scales
appears to be completely dominated by the so-called bottleneck effect,
\ie the pile up of kinetic energy close to the dissipation range due
to the partial suppression of the energy cascade by numerical
viscosity. An inertial range is not recovered until the point where
high resolution $\sim 512^3$, which would be difficult to
realize in global simulations, is reached. We discuss the consequences
for CCSN simulations.
\end{abstract}

\end{abstractbox}

\end{fmbox}
\end{frontmatter}

\section{Introduction}
Despite decades of studies and compelling evidence that a significant
fraction \cite{clausen:15} of stars with initial masses in excess of $\sim$8
solar masses explode as \ac{CCSN} at the end of their evolution, the exact
details of the explosion mechanism are still uncertain \cite{woosley:05nature,
janka:12b, burrows:13a, foglizzo:15}. Current state-of-the art 3D simulations
either fail to explode or have explosion energies that fall short of the
observed energies by factors of a few for most of the progenitor mass
range \cite{janka:12a, burrows:13a, foglizzo:15}.

The dynamics at the center of a star undergoing core collapse is shaped by a
delicate balance between competing effects where all of the known forces:
gravity, electromagnetism, weak and strong interactions, are important. The task
of modeling these systems is made particularly challenging by the fact that the
generation of the  asymptotic explosion energies, although enormous ($\sim
10^{44} \mathrm{J}$), requires a rather subtle, percent-level imbalance between
non-linear processes over many dynamical times.

The flow of plasma in the core of a star going supernova is known to be unstable
to convection \cite{herant:95, bhf:95, janka:96, foglizzo:06} and/or to another
large scale instability known as standing accretion shock instability
\cite{blondin:03, foglizzo:07}. In any case, given the very large Reynolds
numbers, as large as $\sim 10^{17}$ in the region of interest
\cite{abdikamalov:14b} (the so-called gain region, where neutrino heating
dominates over neutrino cooling), it is expected that the resulting flow will be
fully turbulent. It has been suggested \cite{murphy:13, couch:15a} recently that
turbulence and, in particular, turbulent pressure could tip the balance of the
forces in favor of explosion. In this respect, anisotropy is of key importance,
because it results in an effective radial pressure support with adiabatic index
$\gamma_{\mathrm{turb}} = 2$, much larger than that of thermal (radiation)
pressure ($\gamma_{\mathrm{th}} \simeq 4/3$). This means that turbulent kinetic
energy is a much more valuable source of radial pressure support than thermal
energy (see Appendix \ref{sec:turbulent.pressure}).

All of the current numerical simulations employ the \ac{ILES}
paradigm \cite{garnier:09, grinstein:11} (also known as monotone integrated LES
(MILES)) of exploiting the dissipative nature of \ac{HRSC} methods as an
implicit turbulence model.  However, the combination of the use of rather
dissipative schemes and the relatively low spatial resolution that can be
achieved in global simulations is such that the fidelity with which turbulence
is captured is questionable \cite{abdikamalov:14b}.

To be useful in the context of \ac{CCSN} simulations, an \ac{ILES} should, at
the very least, account for the right rate of decay of the kinetic energy at the
largest scales while avoiding unphysical pile up of energy at smaller scales.
Unfortunately, all of the current simulations seem to be strongly dominated by
the so-called bottleneck effect \cite{abdikamalov:14b}, which corresponds to an
inefficient energy transfer across intermediate scales due to the viscous
suppression of non-linear interaction with smaller scales \cite{yakhot:93,
she:93, falkovich:94, verma:07, frisch:08}.  Current global simulations achieve
resolutions, in the turbulent region, comparable to those of $30^3 - 70^3$
lattices in periodic domains \cite{couch:14a,couch:15a,abdikamalov:14b}. At these
resolutions, almost all of the dynamical range of the simulations can be
expected to be directly affected by numerical viscosity \cite{sytine:00}. The
fidelity with which turbulence is captured in these simulations will then depend
on the degree with which the numerical truncation error approximates an LES
closure.

In this respect, it has been shown by \cite{garnier:99} and \cite{johnsen:10}
that many \ac{HRSC} methods can be too dissipative to yield a faithful
description of turbulence at low resolutions. These studies, however, considered
a different regime, decaying isotropic turbulence, while turbulence in a
core-collapse supernova, as well as in many other astrophysical settings, is
often strongly anisotropic \cite{arnett:09, murphy:13, couch:15a} as rotational
invariance is broken by gravity. \cite{garnier:99} and \cite{johnsen:10} also
considered different numerical schemes with respect to those used in supernova
simulations. Both of these aspects can, in principle, be important.  First of
all, strong anisotropies could potentially influence the turbulence dynamics at
the level of the energy cascade and of the dissipation \cite{casciola:07}.
Secondly, some of the schemes used in computational astrophysics, such as the
\ac{PPM} \cite{colella:84} as well as some of the MUSCL \cite{toro:99} schemes,
have been shown, differently from some of the methods considered by
 \cite{garnier:99} and \cite{johnsen:10}, to be well suited for
\ac{ILES} \cite{schmidt:06, thornber:07}.

The aim of this work is to fill the gap between existing theoretical studies and
the particular applications of our interest. To this end we use a publicly
available code, FLASH \cite{fryxell:00, dubey:09, lee:14}, which is widely used
in the computational astrophysics community, and perform a series of simulations
of turbulence in a regime relevant for core-collapse supernovae: driven at large
scale, with large anisotropies and mildly compressible. We use five different
numerical setups and, for each, several resolutions in the range from $64^3$ to
$512^3$ in a periodic domain. We study in detail the way in which the energy
cascade across different scales is represented by our \ac{ILES} and we discuss
the use of local or lower dimensional diagnostics that can be used to assess the
quality of a global simulation in a complex geometry where 3D spectra are not
readily available.

The rest of this paper is organized as follows. First, in Section
\ref{sec:methods}, we discuss the exact setup of our simulations and the
diagnostic quantities used in our analysis. Then, in Section
\ref{sec:results.basic}, we discuss the basic characteristics of the flow
realized in our simulations. In Section \ref{sec:results.energy}, we present a
detailed analysis of the way in which the energy cascade is captured by the
different schemes at different scales. In particular, we quantify the accuracy
with which different methods capture the decay rate of energy from the largest
scales and the way in which energy is distributed across scales. We discuss the
role of anisotropies in the context of the $4/5-$law, a fundamental exact
relation for isotropic and incompressible turbulence relating the statistics of
velocity fluctuations with the energy dissipation rate (see Section
\ref{sec:methods.structure}), in Section \ref{sec:results.45law}. We explore the
use of the 2D, transverse, energy spectrum as a diagnostic for 3D simulations in
Section \ref{sec:results.transverse}.  Finally, we present a brief summary of
our main findings, as well as a discussion of their implications for \ac{CCSN}
simulations in Section \ref{sec:conclusions}. Appendix
\ref{sec:turbulent.pressure}, contains some supplemental background material on
the role of turbulence in the explosion mechanism of \ac{CCSN}.

\section{Methods}
\label{sec:methods}
\subsection{Numerical methods}
We consider a compressible fluid with a prescribed acceleration, $\mathbf{a}$,
in a unit-box with periodic boundary conditions. The code that we employ for
these simulations, FLASH, solves the gas-dynamics equations in conservation
form. In particular we evolve the continuity equation
\begin{equation}\label{eq:continuity}
  \partial_t \rho + \nabla \cdot ( \rho\, \vv ) = 0
\end{equation}
and the momentum equation
\begin{equation}\label{eq:momentum}
  \partial_t (\rho\,\vv) +
    \nabla\cdot(\rho\,\vv\otimes\vv + p\,\mathbf{I}) = \rho\, \mathbf{a}\,.
\end{equation}
These equations are closed with a simple isentropic equation of state,
\begin{equation}\label{eq:eos}
  p = \rho^{4/3}\,,
\end{equation}
that can be considered as a rough description of a gas dominated by radiation
pressure. Since the equation of state ensures an adiabatic evolution we do not
need to solve the energy equation as equations \eqref{eq:continuity},
\eqref{eq:momentum} and \eqref{eq:eos} suffice to fully describe the flow.

Equations \eqref{eq:continuity} \& \eqref{eq:momentum} are solved using the
directionally-unsplit hydrodynamics solver of the open-source FLASH simulation
framework.  FLASH implements the corner transport upwind method
\cite{colella:90} for fully directionally-unsplit evolution of the Euler
equations \cite{lee:09,lee:13}.  FLASH includes several options for the order of
spatial reconstruction \cite{lee:14}, including 2$^{\rm nd}$-order TVD
\cite{toro:99}, 3$^{\rm rd}$-order PPM \cite{colella:84}, and 5$^{\rm th}$-order
WENOZ \cite{borges:08}. Fluxes are computed at 2$^{\rm nd}$ order accuracy using
one of a number of approximate Riemann solvers included in FLASH, such as HLLE
\cite{einfeldt:88} and HLLC \cite{toro:94}. Second-order accuracy in time is
achieved via a characteristic tracing evolution of the Riemann solver input
states to the time step midpoint \cite{colella:84}. We remark that, in
accordance with the \ac{ILES}, paradigm, we do not include any additional
sub-grid scale model, but relied on the implicit turbulent closure built in the
numerical schemes we use for the integration of the hydrodynamics equation.

All of our simulations start with the fluid at rest $\rho = 1$,
$\mathbf{v}=0$. Turbulence is driven using the stirring module of FLASH. This
module uses the Ornstein-Uhlenbeck process \cite{uhlenbeck:30} to generate
stirring modes in Fourier space. This yields an acceleration field which
smoothly decorrelates \cite{eswaran:88} over a timescale $T_s$. The FLASH
implementation permits the use of any arbitrary combination of solenoidal and
compressive modes \cite{federrath:10}. For our runs, we set $T_s = 0.5$, we use
only solenoidal forcing and we restrict the accelerating field to be nonzero
only in the first four Fourier modes. This forcing is designed to mimic the
influence of some larger scale weakly compressible flow and, for this
reason, it does not include any compressible component. This is a reasonable
approximation for low Mach number convection which is well described by the
anelastic approximation, \eg \cite{verhoeven:15}. In the \ac{CCSN}
context, simulations show that the turbulence is highly anisotropic, being
roughly twice as strong in the radial direction as either tangential
direction \cite{murphy:11, murphy:13, handy:14, couch:15a} since it is driven by
buoyancy due to a negative radial entropy gradient. In order to emulate this
behavior, the accelerating field in the $x-$direction (which is going to play
the role of the radial direction) is scaled by a constant factor (before the
solenoidal projection of the acceleration field) such that $R_{xx} \simeq 2
R_{yy} \simeq 2 R_{zz}$, where
\begin{equation}\label{eq:reynolds.stresses}
  R_{ij} = \langle \rho\, v_i\, v_j \rangle\,,
\end{equation}
is the Reynolds stress tensor (to simplify the notation we considered
a frame in which $\langle\rho\vv\rangle = 0$) and
$\langle\cdot\rangle$ denotes an ensemble average. Finally, the
overall strength of the stirring is tuned to achieve a RMS Mach number
of $\simeq 0.35$, which is typically observed in realistic \ac{CCSN}
simulations \cite{couch:13d,mueller:15}.

\subsection{Energy transfer equations}
In order to study the cascade of the specific kinetic energy (which we will
refer to simply as ``kinetic energy'' or ``energy'' in the following),
$|\vv|^2/2$, we will consider an energy budget equation across different scales,
analogous to the one commonly employed in the study of incompressible, isotropic
turbulence, \eg \cite{frisch:96}. In particular, we consider the momentum
equation \eqref{eq:momentum} in non-conservation form,
\begin{equation}\label{eq:velocity}
  \partial_t \vv + (\vv\cdot\nabla)\,\vv = - \VV\,\nabla p + \aa\,,
\end{equation}
where $\VV = 1/\rho$ is the specific volume of the gas.

We can use equation \eqref{eq:velocity} to derive an evolution equation for the
Fourier transform of the velocity
\begin{equation}\label{eq:fourier.transform}
  \hat{\vv}(\kk) = \int_{\mathbb{R}^3} e^{-2\pi\ii \kk\cdot\xx}
    \,\vv(\xx)\,\dd^3 \xx\,.
\end{equation}
Transforming both sides of equation \eqref{eq:velocity} we obtain
\begin{equation}\label{eq:velocity.transform}
  \partial_t \hat{\vv} + \hat{\vv} \ast 2 \pi \ii \kk \otimes \hat{\vv} =
    - \hat{\VV} \ast 2 \pi \ii \kk\, \hat{p} + \hat{\aa}\,,
\end{equation}
where $\ast$ denotes the convolution operator, \ie
\begin{equation}
  [f \ast g](\kk) = \int_{\mathbb{R}^3} f(\qq)\, g(\kk-\qq)\, \dd^3\qq\,.
\end{equation}
If we multiply both sides of equation \eqref{eq:velocity.transform} by
$\hat{\vv}^\ast$ and take the real part, we obtain an equation for the 3D energy
spectrum
\begin{equation}\label{eq:energy.3d}
  \partial_t E(\kk) = T(\kk) + C(\kk) + \epsilon(\kk)\,,
\end{equation}
where
\begin{align}
  \label{eq:energy.spectrum}
  E(\kk) &= \frac{1}{2}\, \hat{\vv}\cdot\hat{\vv}^\ast\,, \\
  T(\kk) &= - 2\pi\, \Re\, (\hat{\vv} \ast \ii\kk\otimes\hat{\vv})
            \cdot \hat{\vv}^\ast\,, \\
  C(\kk) &= - 2\pi\, \Re\, (\hat{\VV}\ast\ii\kk\, p) \cdot \hat{\vv}^\ast\,, \\
  \epsilon(\kk) &= \Re\, \hat{\aa}\cdot\hat{\vv}^\ast\,.
\end{align}
Here $E$ is the energy spectrum (the velocity \ac{PSD}) and $T$ is the same
transfer term as in the classical incompressible equations and $\epsilon$ is the
energy injection rate. The $C$ term vanishes in the incompressible limit and
represents the interaction between kinetic and acoustic modes. In practice, in
our models, $C$ is found to be at least one order of magnitude smaller than $T$
at all scales and it is thus negligible. In any case, we retain $C$ in the
analysis below.

For each of the spectral quantities, $S$, being $E, T, C$ or $\epsilon$, we
define the integrated spectrum, $S(k)$, as
\begin{equation}\label{eq:sph.integral}
  S(k) = \int_{\mathbb{R}^3} S(\kk)\, \delta(|\kk| - k)\, \dd \kk\,,
\end{equation}
$\delta(\cdot)$ being the Dirac delta function.

Integrating equation \eqref{eq:energy.3d}, we obtain the following
one-dimensional energy balance equation
\begin{equation}\label{eq:energy.integrated}
  \partial_t E(k) = T(k) + C(k) + \epsilon(k)\,.
\end{equation}
This can also be written in terms of the energy flux across scales,
\begin{equation}\label{eq:energy.flux}
  \Pi(k) = - \int_0^k T(\xi)\, \dd \xi\,,
\end{equation}
as
\begin{equation}\label{eq:energy.1d}
  \partial_t E(k) + \partial_k \Pi(k) = C(k) + \epsilon(k)\,.
\end{equation}
Notice that \emph{we did not assume isotropy in any of the above}.

Equation \eqref{eq:energy.integrated} is derived in the inviscid limit. In
practice, our evolution method introduces dissipation in the form of ``numerical
viscosity''. This can be quantified in terms of the residual
\begin{equation}
  R(k) = \partial_t E(k) - T(k) - C(k) - \epsilon(k)\,.
\end{equation}
This can be used to define a wave number dependent numerical viscosity:
\begin{equation}\label{eq:effective.viscosity}
  \nu(k) = - \frac{1}{2} \frac{R(k)}{k^2\, E(k)}\,.
\end{equation}
We remark that $\nu$ does not, in general, correspond to a classical shear or
bulk viscosity, but can nevertheless be interpreted as a relative measure of the
dissipation acting at different wave numbers (see, \eg \cite{fureby:99, aspden:09,
zhou:14} for alternative approaches).

In practice, since we will be working in the stationary case, after having taken
the appropriate time averages, $R(k)$ reduces to
\begin{equation}
  R(k) =  - T(k) - C(k) - \epsilon(k)\,.
\end{equation}

Finally, since we are working in a periodic domain, which we take of size $L_x =
L_y = L_z = 1$, all of the spectra are quantized and non-trivial only for $k_x,
k_y$ and $k_z$ integers. Furthermore, all of the integrals in wave number space
reduce to summations. Integrals over spherical shells are transformed to
weighted sums following \cite{eswaran:88}:
\begin{equation}
  E(k) = \frac{4\pi k^2}{N_k} \sum_{k-1/2 < |\kk| \leq k+1/2} E(\kk)\,,
\end{equation}
where $N_k$ is the number of discrete wave-numbers in the shell $k-1/2
< |\boldsymbol{k}| \leq k+1/2$.

\subsection{Structure functions}
\label{sec:methods.structure}
The energy spectrum and its sources/fluxes give a comprehensive picture of the
energy cascade and can be used to assess the level of convergence of the
simulation. Unfortunately, 3D energy spectra and fluxes are not easily
accessible in calculations in complex domains and/or with inhomogeneous
turbulence. In these cases, local quantities in the physical domain are more
easily extracted and analyzed. Hence, one of the goals of this work is to
validate the use of indirect measures of convergence of \ac{ILES}. Among these
quantities, the structure functions of the velocity appear to be natural
candidates for study.

We define the velocity increments
\begin{equation}\label{eq:delta.vel.para}
  \delta v(\xx, \rr) = \big[\vv(\xx + \rr) - \vv(\xx)\big]\cdot\frac{\rr}{r}
\end{equation}
and study the quantities
\begin{equation}
  \label{eq:structure.functions}
  S_p(r) = \langle \delta v^p \rangle_{j=0}\,,
\end{equation}
where, $\langle\cdot\rangle_{j=0}$ denotes an ensemble average as well as a mean
over all of the angles between $\vv$ and $\rr$ (in other words we are looking at
the $j=0$ component of the $\mathrm{SO}(3)$ decomposition of the structure
functions \cite{biferale:05}). In the case of homogeneous turbulence $S_p$
does not depend on $\xx$ and is thus a function of only the separation $r$.

The most important relation involving the structure functions is the so-called
$4/5-$law, which relates the third order structure function, $S_3(r)$, with the
mean energy dissipation rate,
\begin{equation}\label{eq:dissipation.rate}
  \langle\epsilon\rangle = \int_0^\infty \epsilon(k)\, \dd k\,,
\end{equation}
and states that, for incompressible, homogeneous and isotropic
turbulence \cite{frisch:96}:
\begin{equation}\label{eq:45law}
  S_3(r) = - \frac{4}{5}\, \langle\epsilon\rangle\, r\,.
\end{equation}
Equation \eqref{eq:45law} can be derived from the Navier-Stokes equation for
fully-developed, incompressible, homogeneous and isotropic turbulence and it is
one of the few exact relations in the theory of turbulence \cite{frisch:96}. In
the anisotropic or compressible case, however, equation \eqref{eq:45law} is not
strictly valid and could be violated in the data. As we show in Section
\ref{sec:results.45law}, we find equation \eqref{eq:45law} to be very well
satisfied by our data, suggesting that the 3$^{\rm rd}$ order structure function
can be a very useful diagnostic in global simulations.

\subsection{Transverse energy spectrum}
Another alternative to the analysis of 3D spectra, which has been adopted by
several authors in the core-collapse supernova context \cite{dolence:13,
couch:14a, handy:14, abdikamalov:14b}, is the use of 2D spectra computed using a
spherical harmonics expansion of the velocity field tangential to one or more
spherical shells in the simulation. Analogously, we emulate this by looking at
quantities in the $y-z$ plane and we define the 2D spectra
\begin{equation}\label{eq:energy.spectrum.2d}
  E_\perp(k_\perp) = \frac{1}{2} \int_{\mathbb{R}^2} \tilde{\vv}_\perp \cdot
  \tilde{\vv}^\ast_\perp\, \delta\left(\sqrt{k_y^2 + k_z^2} - k_\perp\right)\,
  \dd k_y\, \dd k_z\,,
\end{equation}
where $\mathbf{v}_\perp$ is the projection of the velocity perpendicular to the
$x-$direction and we introduced the partial Fourier transform of $\vv_\perp$:
\begin{equation}\label{eq:partial.fourier.transform}
\begin{split}
  \tilde{\vv}_\perp(k_y, k_z) = & \lim_{L_x\to+\infty} \frac{1}{L_x}
  \int^{L_x/2}_{-L_x/2} \dd x \\ & \int_{\mathbb{R}^2} e^{-2\pi\ii (k_y y + k_z z)}\,
  \mathbf{v}_\perp(x,y,z)\, \dd y\, \dd z\,.
\end{split}
\end{equation}
In the limit of infinite Reynolds number / resolution, the 2D spectrum is
expected to have the same asymptotic behavior as the 3D spectrum, however it is
a-priori unclear if $E_\perp$ is a good proxy for $E$ at finite resolution.  For
this reason we find it useful to investigate this here.

As was the case for the 3D spectra, also here the spectrum is non-trivial only
for integer $k_y$ and $k_z$, when periodicity is taken into account. The
integral in equation \eqref{eq:energy.spectrum.2d} is treated analogously to the
integral in the equation \eqref{eq:sph.integral} for the 3D case, while the
average in the $x-$direction in equation \eqref{eq:partial.fourier.transform} is
converted to an average over the $x-$extent of the simulation box.

\section{Basic flow properties}
\label{sec:results.basic}
We employ the finite-volume \ac{HRSC} (Godunov) approach in which physical
states are reconstructed at inter-cell boundaries and local Riemann problems are
solved to compute the physical inter cell fluxes.  In particular, we perform
five groups of simulations using different numerical methods. Each group is
labeled using the name of the reconstruction algorithm and of the Riemann
solver. For instance \texttt{TVD\_HLLE}, denotes a group of simulations done
using TVD reconstruction and HLLE Riemann solver. Single simulations are labeled
using their resolution so that, for instance, \texttt{TVD\_HLLE\_N128}, denotes
the \texttt{TVD\_HLLE} run done using a $128^3$ grid. For all of the runs the
timestep is chosen to have a CFL, \ie $c \Delta t / \Delta x$, of $0.4$, $c$
being the maximum characteristics speed, with the exception of the
\texttt{PPM\_HLLC\_CFL0.8} runs where we set the CFL to $0.8$. For the TVD runs
we use the \ac{MC} slope limiter \cite{toro:99}. The runs with PPM use the
original flattening and artificial viscosity prescriptions from
\cite{colella:84}. The artificial viscosity coefficient is $0.1$. We remark
that the use of the artificial viscosity for PPM is not really necessary in this
regime \cite{porter:94}, however our goal is not to perform a study of the
turbulent dynamics, but to assess how each numerical method performs when used
under the same condition as in a real \ac{CCSN} simulation where strong shocks
need to be handled in some parts of the domain.

\begin{figure*}
\begin{center}
  \includegraphics[scale=0.8]{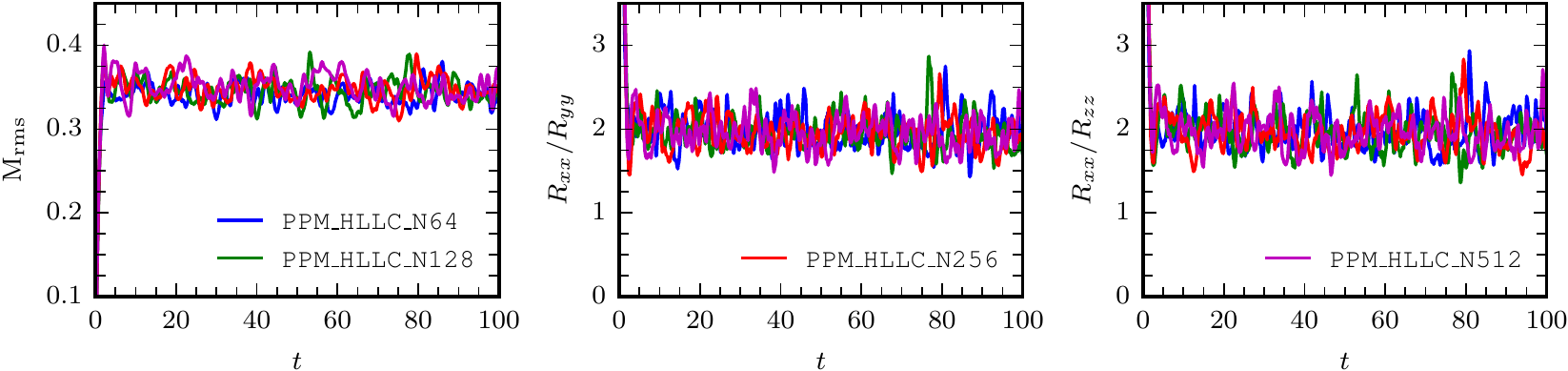}
  \caption{Time evolution of the diagnostic quantities for the fiducial set of
  runs \texttt{PPM\_HLLC} with different resolutions. The left panel shows the
  \ac{RMS} Mach number, while the middle and left panels show, respectively, the
  ratios $R_{xx}/R_{yy}$ and $R_{xx}/R_{zz}$, $R$ being the Reynolds stress
  tensor (equation \ref{eq:reynolds.stresses}). Since the $x-$direction is the
  anisotropic direction (it would play the role of the radial direction in a
  \ac{CCSN}) the ratios $R_{xx}/R_{yy}$ and $R_{xx}/R_{zz}$, offer a global
  measure of the anisotropy of the flow at the largest scale. All of the
  quantities appear to have reached stationarity after time $t \gtrsim 3$ and
  oscillate around their target values. All resolutions produce the same
  qualitative behavior.  \label{fig:time.evolution}}
\end{center}
\end{figure*}

For each group of simulations we run four resolutions: $64^3, 128^3, 256^3$ and
$512^3$. The RMS velocity in all of the runs is $v_{\mathrm{rms}} \simeq 0.4$,
giving an eddy turnover time $\tau = 1/v_{\mathrm{rms}} \simeq 2.5$.  All of the
simulations are run until time $t = 100$ ($\simeq 40$ eddy turnover times). The
time evolution of a few relevant diagnostics is shown in Figure
\ref{fig:time.evolution} for our fiducial group of runs (\texttt{PPM\_HLLC}) at
different resolutions. We can see how the flow is accelerated from rest and
quickly reaches a steady, fully turbulent, state. In all cases, steady state is
reached after $t\gtrsim 3$ ($\sim 1$ turnover time) and the diagnostics are
insensitive to the resolution. The results for the other runs (not shown) are
very similar to the ones of \texttt{PPM\_HLLC} as they all achieve very similar
RMS Mach numbers and Reynolds stresses. All of the analysis shown in the
rest of the paper are performed using $380$ 3D snapshots (evenly spaced in time)
of the data in the interval $5 \leq t \leq 100$.

\begin{figure*}
  \centering
  \includegraphics[scale=0.8]{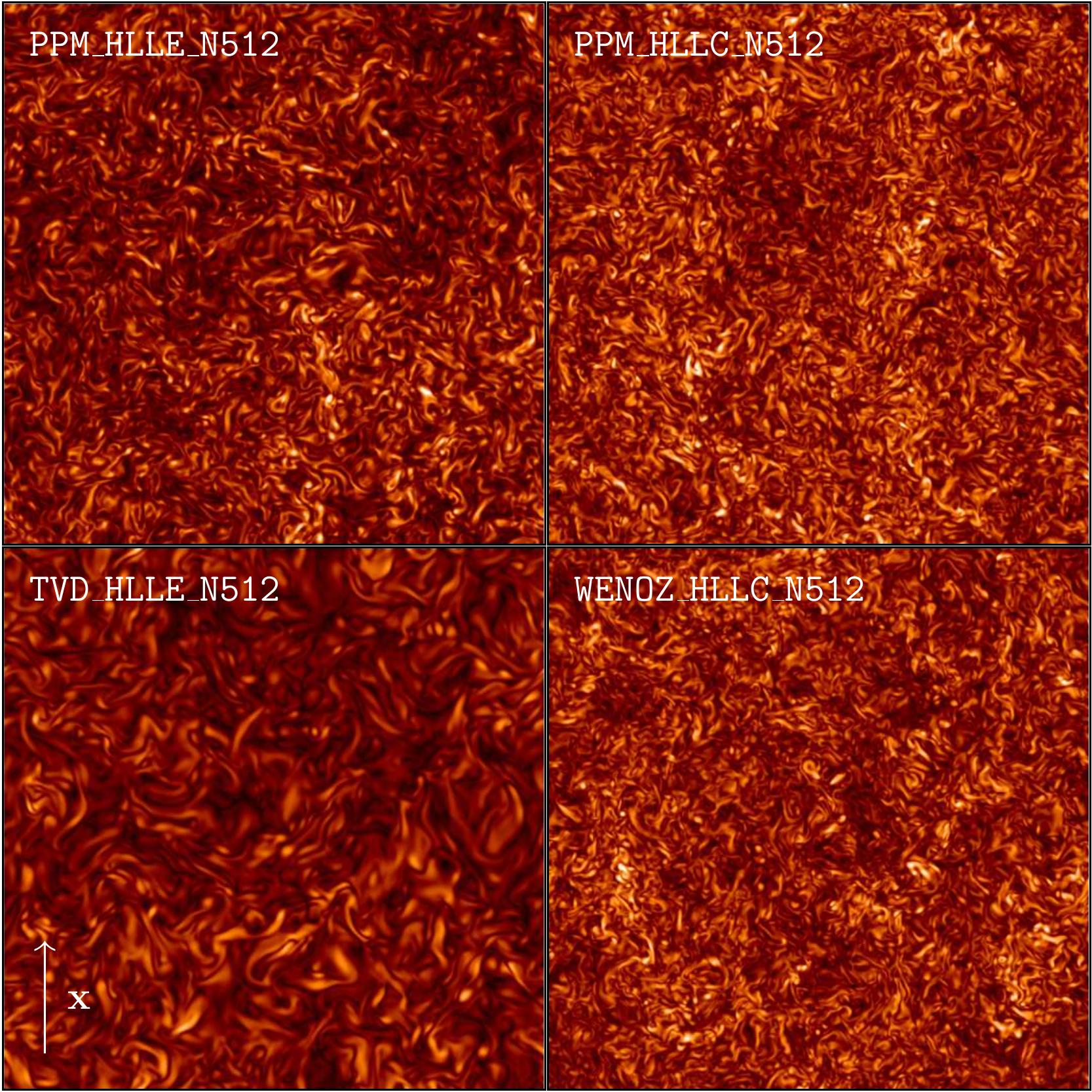}
  \caption{Square root of the magnitude of the vorticity,
  $\sqrt{|\nabla\times\mathbf{v}|}$, for four of the simulations with 512$^3$
  resolution in a slice through the middle of the $x$--$z$ plane at the final
  time of the simulations ($t=100$).  The panels show simulations using
  \texttt{PPM\_HLLE\_N512}, \texttt{PPM\_HLLC\_N512},
  \texttt{WENOZ\_HLLC\_N512}, and \texttt{TVD\_HLLC\_N512} clockwise from the
  top left. The direction of the anisotropic driving is up in these figures. The
  colorcode goes linearly from $0$ (no vorticity; dark colors) to $15$ (light
  colors) and it is the same for all panels.}
  \label{fig:vorticity}
\end{figure*}

A first, qualitative, comparison between the different methods can be done by
looking at their visualizations. In particular, in Figure \ref{fig:vorticity},
we show a visualization of the magnitude of the vorticity in the $x$--$z$ plane
for four of the five schemes (excluding \texttt{PPM\_HLLC\_CFL0.8}) at the
highest resolution ($512^3$). The data is taken at the final time ($t = 100$).
As it can be seen from the figure, all of the simulations show the presence of
thin, elongated, regions of high vorticity, as typically seen in \ac{DNS} of
homogeneous turbulent flows \cite{vincent:91, ishihara:07}. However, the width
and the intensity of the vorticity at these smaller scales depend crucially on
the numerical scheme. Methods with small intrinsic numerical viscosity, such as
\texttt{PPM\_HLLC} and \texttt{WENOZ\_HLLC}, present smaller structures and more
intermittent vorticity fields with respect to more dissipative methods, such as
\texttt{PPM\_HLLE} and \texttt{TVD\_HLLE}.


\section{The energy cascade}
\label{sec:results.energy}
In this section we focus our analysis on the accuracy with which the energy
cascade is captured by our \ac{ILES} runs. First, we focus on the largest scales
of the simulation with the goal of quantifying the accuracy in the decay rate of
the energy as a function of the resolution for the different methods.  Next, we
will look at the energy distribution at smaller scales where, in resolved
simulations, the inertial range starts. Finally, we will look at the dynamics in
the dissipation region and summarize.

\subsection{Energy decay rate}
\label{sec:results.energy.decay}

\begin{figure*}
\begin{center}
  \includegraphics[scale=0.8]{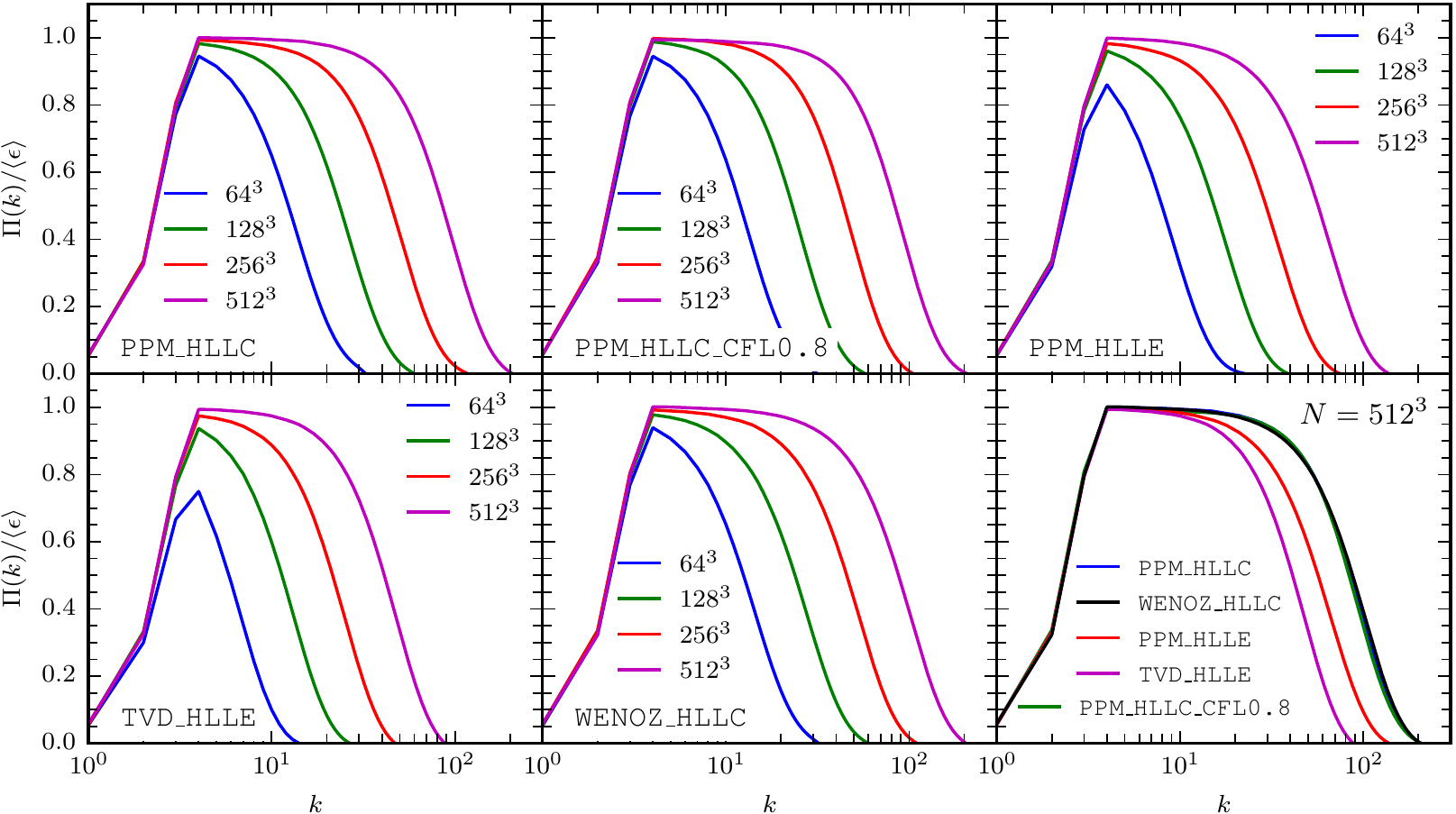}
  \caption{Energy flux, as defined by equation \eqref{eq:energy.flux}, obtained
  with different numerical methods and resolutions. The energy flux is shown
  normalized to the average dissipation rate given by equation
  \eqref{eq:dissipation.rate}. From left to right and from top to bottom we show
  the results obtained with \texttt{PPM\_HLLC}, \texttt{PPM\_HLLC\_CFL0.8},
  \texttt{PPM\_HLLE}, \texttt{TVD\_HLLE} and \texttt{WENOZ\_HLLC}. The bottom
  right panel show a comparison of all of the methods at $512^3$. All of the
  schemes show a good level of accuracy in the energy flux from the largest
  scales, with errors smaller than a few \% already at low resolutions. The
  differences between the schemes become more marked at large wave numbers where
  the numerical dissipation starts to interfere with the energy cascade.
  \label{fig:energy.flux}}
\end{center}
\end{figure*}

In the limit of very large Reynolds number it is assumed, in standard
turbulence phenomenology \cite{frisch:96}, that there exists a range of
wave numbers (the inertial range) where energy injection and dissipation can be
neglected in equation \eqref{eq:energy.1d}. In this range we can write
(compressible effects are negligible in our simulations):
\begin{equation}
  \partial_t E(k) + \partial_k \Pi(k) \simeq 0\,,
\end{equation}
so that stationarity requires $\Pi(k) \simeq \mathrm{const}$. In particular, since
energy is conserved, one finds $\Pi(k) \simeq \langle\epsilon\rangle$. This means
that, in the limit of large Reynolds numbers, the energy decay rate depends only
on the macroscopic properties of the flow (and in particular not on the nature
of the viscosity), a fact that has also been verified
numerically \cite{kaneda:03}. The significance of this property and its
importance for the modeling of turbulence cannot be overstated.

In the context of \ac{CCSN} simulations this means that the large scale kinetic
energy, a crucial quantity for the dynamics of the explosion \cite{couch:15a},
can be faithfully captured even with simulations achieving modest Reynolds
numbers.

For an \ac{ILES}, a basic requirement, then, is that a sufficiently high
resolution should be achieved to correctly represent the energy cascade at the
largest scales. What qualifies as a sufficiently high resolution is of course
dependent on the details of the closure built into the scheme (and on the
accuracy required for the particular application). To quantify this, we can
estimate the level of accuracy that can be reached at any given resolution,
using our local simulations. In particular, we can study directly the energy
flux across scales, defined by equation \eqref{eq:energy.flux}. This is shown in
Figure \ref{fig:energy.flux} for all of the different runs,.

As discussed before, we expect that $\Pi(k) \simeq \langle\epsilon\rangle$ over
an extended region in Fourier space should be a direct indication that a
simulation has been able to recover an inertial range. Perhaps not
surprisingly, in light of previous results \cite{sytine:00}, we find that
regions where $\Pi\simeq\langle\epsilon\rangle$ as wide as a few wave numbers $4
\lesssim k \lesssim 10$ only appear at the highest resolutions (we will discuss
the inertial range in more detail in Section \ref{sec:results.energy.spectra}).
However, the amount of energy decaying from the largest scales reaches an
asymptotic value much quicker than that implying that the total kinetic energy
budget at the largest scales is well resolved even at modest resolutions.

\begin{figure}
\begin{center}
  \includegraphics[scale=0.8]{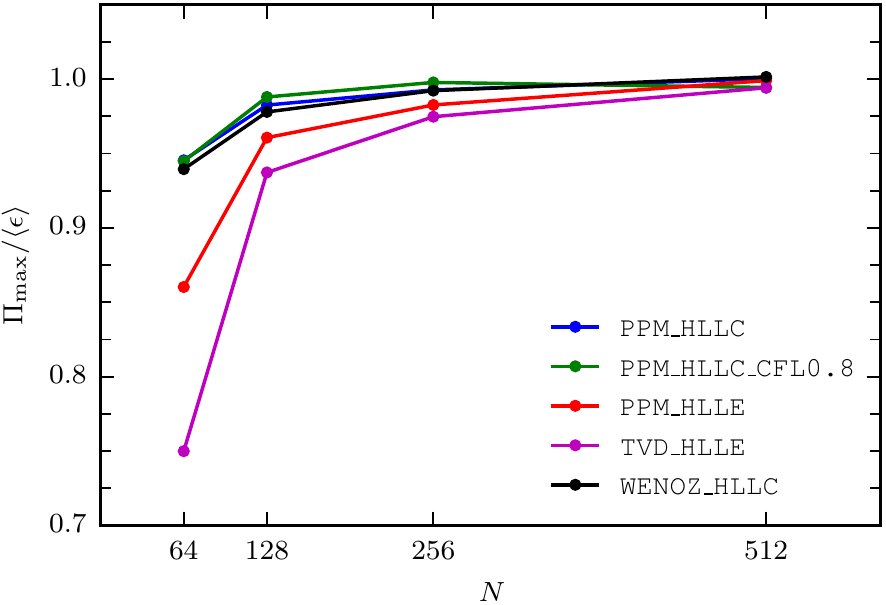}
  \caption{Dissipation rate of the energy at the largest scales due to the
  turbulent cascade (not including direct dissipation by the numerical
  viscosity) as a function of resolution and for all of the schemes. The
  dissipation rate is normalized so as to be $1$ in the limit of large Reynolds
  numbers / resolution. At $128^3$ points all of the schemes show an error of
  less than $10\%$, with the HLLC schemes already close to the $2\%$ level.}
  \label{fig:dissipation.rate}
\end{center}
\end{figure}

We can make a more quantitative statement concerning the energy decay rate by
looking at the peak of the energy flux as a function of resolution, as shown in
Figure \ref{fig:dissipation.rate}. We can see that at $128^3$ points all of the
simulations have a deviation from the asymptotic energy decay rate of less than
$10\%$. The least dissipative methods already have an error close to the $2\%$
level. A comparison between \texttt{PPM\_HLLE} and \texttt{PPM\_HLLC} reveals
the profound impact that the choice of the Riemann solver has even at relatively
large scale (more on the dissipative properties of the different schemes in
Section \ref{sec:results.energy.viscosity}).

\subsection{Energy spectra}
\label{sec:results.energy.spectra}

\begin{figure*}
\begin{center}
  \includegraphics[scale=0.8]{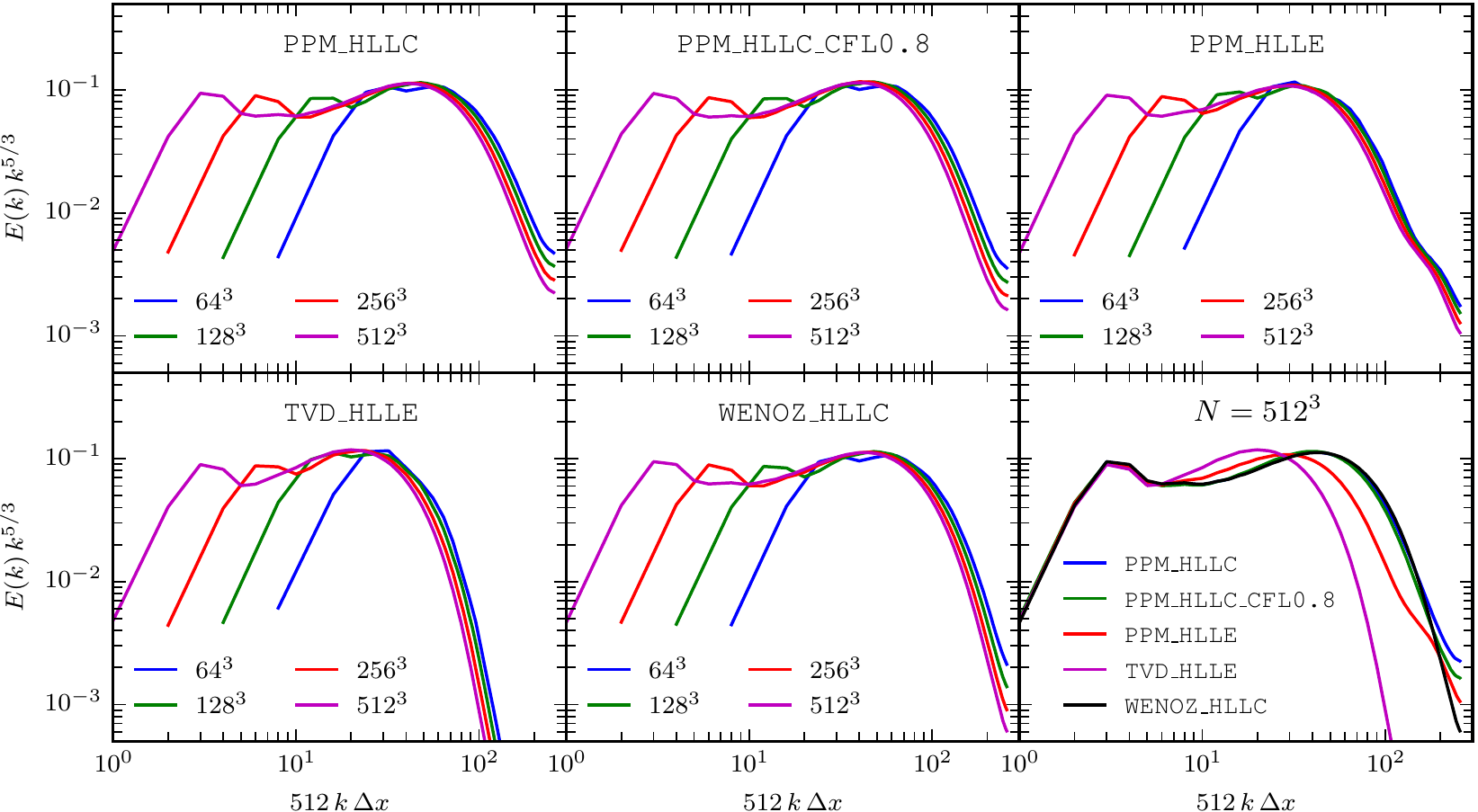}
  \caption{Energy spectra (equation \ref{eq:energy.spectrum}) obtained with
  different numerical methods and resolutions. The energy spectra are
  compensated by a $k^{5/3}$ spectrum, so that any region with Kolmogorov
  scaling should appear roughly flat. Furthermore, the spectra are all plotted
  as a function of the dimensionless wave number $512\,k\,\Delta x$ (the $512$
  factor is introduced to have the dimensionless wave number coincide with the
  dimensional one for the $512^3$ runs). The first five panels show the
  \texttt{PPM\_HLLC} (upper left), \texttt{PPM\_HLLC\_CFL0.8} (upper center),
  \texttt{PPM\_HLLE} (upper right), \texttt{TVD\_HLLE} (lower left) and
  \texttt{WENOZ\_HLLC} (lower center) group of runs. The last panel (lower
  right) shows a comparison of all of the methods at the highest resolution
  ($512^3$). An inertial range seems to be recovered only at the highest
  resolutions (perhaps with the exception of \texttt{TVD\_HLLE} where no
  inertial range is visible). All schemes employing the HLLC Riemann
  solver are in very good agreement.
  \label{fig:energy.spectrum}}
\end{center}
\end{figure*}

Obviously, not all of the dynamics of turbulence can be reduced to the rate at
which kinetic energy decays from the injection scale. The internal dynamics of
the energy cascade, far from the injection scale and far from the dissipation
range, can also play an important role in many applications. To analyze this
aspect we consider in Figure \ref{fig:energy.spectrum} the energy spectrum of
the velocity defined by equation \eqref{eq:energy.spectrum}. The spectra are
compensated by $k^{5/3}$ to highlight regions with Kolmogorov scaling,
which might be expected in the inertial range. Since we want to focus on
quantities that do not depend (or depend weakly) on the nature of the energy
injection at large scale, we show all of the spectra as a function of a
dimensionless wave number, $512\, k\, \Delta x$. The rationale behind this
normalization is that, first of all, we assume the Kolmogorov scale $\eta$ to be
proportional to the grid spacing.  Secondly, the $512$ factor is introduced to
have the dimensionless $k$, $512\, k\, \Delta x$ coincide with the dimensional
one for the highest resolution runs.  With this choice, $512\, k\, \Delta x=512$
corresponds to a wavelength of a single grid point, $512\, k\, \Delta x=256$
corresponds to a wavelength of two grid points and so on.

Looking at any of the groups of runs in Figure \ref{fig:energy.spectrum}, one
can immediately notice that the spectra obtained at different resolutions do not
collapse into a single curve in the dissipation region, as would be required by
Kolmogorov's first similarity hypothesis \cite{frisch:96} (\cf  \cite{gotoh:02}).
This lack of convergence in the dissipation region could be due to the
non-linear viscosity of \ac{HRSC} schemes. This, in turn, could result in an
anomalous scaling of $\eta$ with the grid spacing. Such scaling has been
reported in the past for \ac{ILES}, but it is not very well
understood \cite{aspden:09}. The good agreement between the three different
groups of simulations employing the HLLC Riemann solver seems to support this
hypothesis and suggests that the nonlinear viscosity introduced by the Riemann
solver is an important ingredient in setting this scaling.

Convergence appears to be recovered at larger scales $\gtrsim 8\, \Delta x$
($512\, k\, \Delta x \lesssim 64$), but the spectra appear to be dominated by
the bottleneck effect. This manifests itself as a bump in the compensated
spectra extending from the dissipation range until the end of the inertial
range, for the simulations that show one (\eg  until $512\, k\, \Delta x = 10$
for the HLLC runs), or until the energy injection scale ($512\, k\, \Delta x =
4$), for the simulations that show no or little inertial range
(\texttt{TVD\_HLLE}). The bottleneck effect is a viscous phenomenon which is
also observed in direct numerical simulations.  However, in the present context
where viscosity is of numerical origin, it is at the very least questionable if
a pronounced bottleneck is a desirable feature of the modeling.  In
astrophysical flows, where the Reynolds numbers are typically very large, this
pile up of energy at large scales is unphysical and could affect the
quantitative and qualitative outcome of a simulation \cite{abdikamalov:14b}. A
quantification of the bottleneck effect in terms of the energy budget is
discussed in Section \ref{sec:results.energy.cumulative}.

At even larger scales, an inertial range ($E\sim k^{-5/3}$ and $\Pi \sim
\mathrm{const}$, see Figure \ref{fig:energy.flux}) seems to be recovered by the
least dissipative schemes (PPM and WENOZ with HLLC) in the region $4\lesssim k
\lesssim 10$. \texttt{PPM\_HLLE} and \texttt{TVD\_HLLE} have a more limited
region, a few wave numbers at most, that could be interpreted as being an
inertial range. We note that this resolution is not particularly high in
comparison with state of the art \ac{DNS} \cite{kaneda:03, federrath:13}, but it
would already correspond to an extremely high resolution in global \ac{CCSN}
simulations that typically have $\sim 30 - 70$ zones across the turbulent
region \cite{abdikamalov:14b}.

The overall behavior of the spectra, as obtained by all schemes, is
consistent with Kolmogorov's theory of turbulence. The anisotropic
contributions to the angle-integrated spectra are too small to be
detected in our data.

\subsection{Numerical viscosity}
\label{sec:results.energy.viscosity}

\begin{figure*}
\begin{center}
  \includegraphics[scale=0.8]{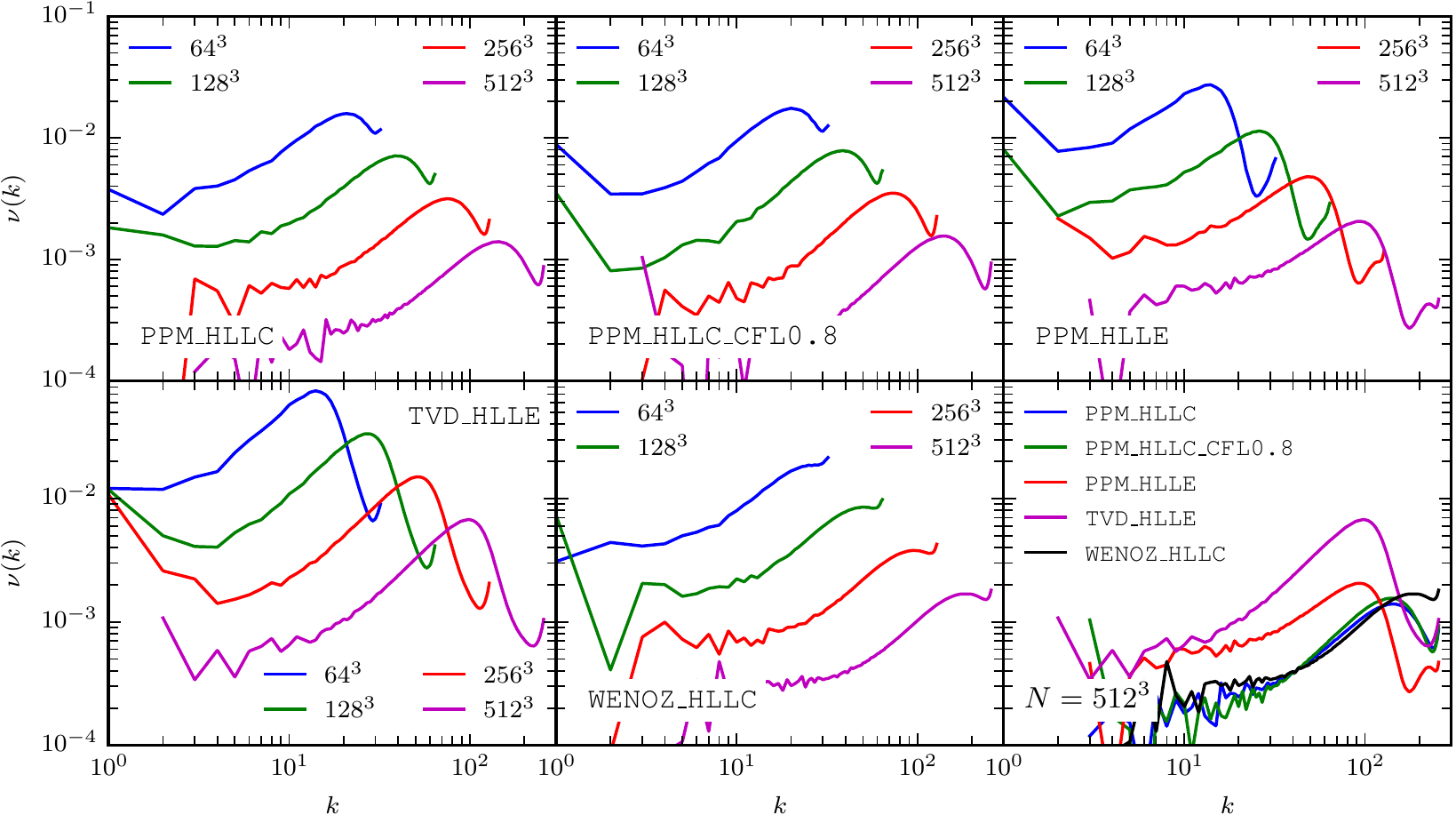}
  \caption{Numerical viscosity as a function of the wave number measured for all
  schemes and resolutions. The numerical viscosity is estimated using the
  procedure outlined in Section \ref{sec:methods} and it is defined by equation
  \eqref{eq:effective.viscosity}. The different panels are, from left to right
  and from top to bottom, the results obtained with \texttt{PPM\_HLLC},
  \texttt{PPM\_HLLC\_CFL0.8}, \texttt{PPM\_HLLE}, \texttt{TVD\_HLLE} and
  \texttt{WENOZ\_HLLC}. The bottom right panel show a comparison of all of the
  methods at $512^3$. The numerical viscosity shows large variations across the
  wave number space. The choice of the Riemann solver plays a role that is at
  least as important as the choice of the reconstruction method in affecting the
  numerical viscosity throughout the entire the
  spectrum.\label{fig:effective.viscosity}}
\end{center}
\end{figure*}

At very small scales ($\sim$ several grid points) the dynamics is dominated by
the numerical viscosity. This can be estimated from the residual of the energy
equation \eqref{eq:energy.1d} or, equivalently, by the effective numerical
viscosity $\nu(k)$ (equation \ref{eq:effective.viscosity}). The latter is shown
in Figure \ref{fig:effective.viscosity} for all schemes and resolutions.

The first thing to notice is that the numerical viscosity provided by all
numerical schemes is not constant, but differs by roughly an order of magnitude
between low and high $k$. Having a wave number dependent viscosity is a
desirable feature expected in any LES model (explicit or otherwise).
Nevertheless, this makes the definition and calculation of the effective
Reynolds number achieved in a simulation ambiguous. Meaningful ways to estimate
it for \ac{ILES} have been proposed \cite{zhou:14} and they can be used to ease
the comparison between different simulations and assess their quality. However,
one has to be very careful while using any quoted ``Reynolds number'' from an
\ac{ILES}, to estimate things like the dynamical range achieved by a simulation,
because the dissipative properties of \ac{ILES} differ considerably from the
ones of the true Navier-Stokes equations.

Two other features can be observed in most of the numerical viscosity profiles.
First, many of them exhibit a sudden reversal at high wave numbers. This is due
to the fact that the numerical viscosity does not behave like a shear viscosity
so that, although the numerical diffusion is strong at those scales, the
numerical viscosity appears small because of a partial decoupling between
vorticity and dissipation. Second, at high resolution and at the largest scales,
the numerical viscosity is close to zero or even slightly negative. The reason
is that the residual of equation \eqref{eq:energy.3d} oscillates around zero and
it is too small to be reliably extracted from our data: a much longer
integration time would be needed to accumulate enough statistics for it.

Finally, a comparison between the numerical viscosity reveals two interesting
effects.  First, by comparing \texttt{PPM\_HLLC} and \texttt{PPM\_HLLE}, we see
that the choice of the Riemann solver affects the viscosity at basically all
scales. Second, if we compare \texttt{PPM\_HLLC}, \texttt{PPM\_HLLC\_CFL0.8} and
\texttt{WENOZ\_HLLC}, we see that doubling the timestep appears to have an
effect comparable to the difference between the PPM and WENOZ reconstructions at
intermediate scales ($40\lesssim k \lesssim 100$).

\subsection{The energy distribution}
\label{sec:results.energy.cumulative}

\begin{figure*}
\begin{center}
  \includegraphics[scale=0.8]{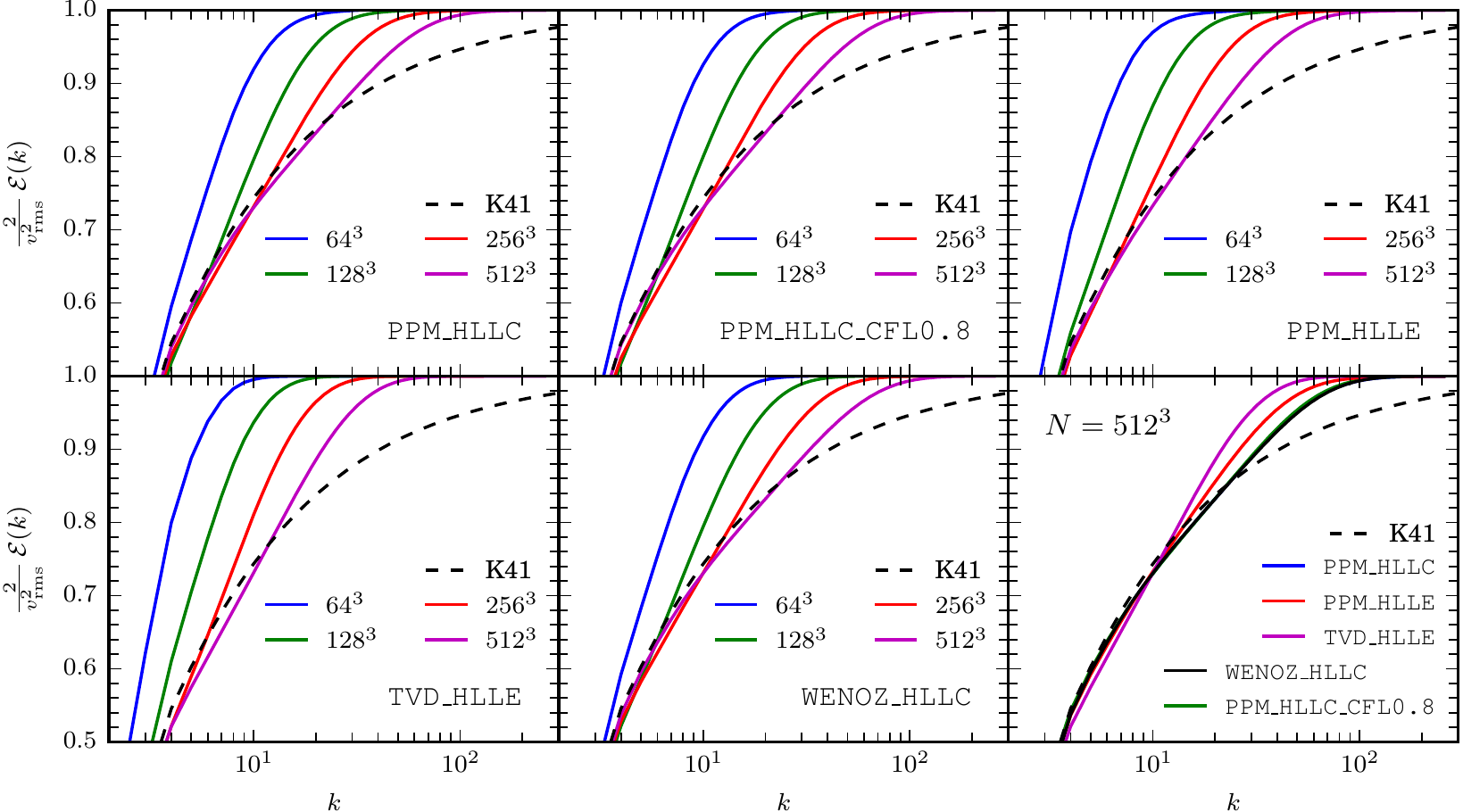}
  \caption{Cumulated energy distribution (equation \ref{eq:cumulative.energy})
  for all methods and resolutions, normalized by a factor $2/v_{\mathrm{rms}}^2$
  to be equal to $1$ for large $k$. As a reference for comparison we also plot
  the asymptotic profile expected from Kolmogorov's theory (equation
  \ref{eq:cumulative.energy.k41}). The different panels are, from left to right
  and from top to bottom, the results obtained with \texttt{PPM\_HLLC},
  \texttt{PPM\_HLLC\_CFL0.8}, \texttt{PPM\_HLLE}, \texttt{TVD\_HLLE} and
  \texttt{WENOZ\_HLLC}. The bottom right panel show a comparison of all of the
  methods at $512^3$. At low resolution all of the schemes show an excess of
  energy at intermediate scales, due to the bottleneck. Only at the highest
  resolution at least, roughly, $80\%$ of the energy is correctly resolved.}
  \label{fig:cumulative.energy}
\end{center}
\end{figure*}

So far we have been concerned with the energy decay rate from the largest
scales, which we have shown to be well captured by the \ac{ILES} (Section
\ref{sec:results.energy.decay}), and with the energy transfer in the inertial
range, which we have seen to be described accurately only at much higher
resolutions (Section \ref{sec:results.energy.spectra}). In a turbulent flow both
of these aspects are important and a good \ac{ILES} should display a
distribution of energy across vortical structures at different scales that is as
close as possible to the asymptotic one. Obviously, there is a limit to the
accuracy that any \ac{ILES} can achieve at a fixed resolution. Here, we make
this statement more quantitative by considering the amount of kinetic energy
that is well resolved by each simulation at a given resolution.

We introduce the cumulative energy spectrum, the integral of the energy
spectrum:
\begin{equation}\label{eq:cumulative.energy}
  \mathcal{E}(k) = \int_0^k E(\xi)\, \dd \xi\,.
\end{equation}
This quantity is plotted in Figure \ref{fig:cumulative.energy}, where it is
normalized by 
\begin{equation}
  \frac{v_{\mathrm{rms}}^2}{2} = \int_0^{+\infty} E(k)\, \dd k
\end{equation}
to obtain the cumulative distribution function of the kinetic energy. As a
reference, we also show the cumulative energy spectrum estimated from
Kolmogorov's theory:
\begin{align}\label{eq:cumulative.energy.k41}
  &\mathcal{E}_{\textrm{K41}}(k) = \int_0^k E_{\textrm{K41}}(\xi)\, \dd \xi\,, \\
  \begin{split}
  &E_{\textrm{K41}}(k) = \\ & \qquad \begin{cases}
    E_{\texttt{PPM\_HLLC\_N512}}(k)\,, & \textrm{if } k \leq 4\,, \\
    E_{\texttt{PPM\_HLLC\_N512}}(4) \left(\frac{k}{4}\right)^{-5/3}, &
      k > 4\,.
  \end{cases}
  \end{split}
\end{align}
We find that as the resolution increases, all schemes appear to be converging to
the predictions of Kolmogorov's theory. The results at finite resolution,
however, are not encouraging: at $64^3$ only $\sim 50\%$ or less of the kinetic
energy is in well resolved structures, while the other $\sim 50\%$ have piled up
at rather large scale, with a cumulative excess of $\sim 20\%$ at the grid
scale, mostly because of the bottleneck effect. At higher resolutions, the
amount of kinetic energy well captured by the \ac{ILES} increases, but at
$512^3$ this is still only about $80\%$ of the energy and there is still a
cumulative excess of over $\sim 5\%$ at the grid scale ($\ell \sim \Delta x$).

\section{The $4/5-$law}
\label{sec:results.45law}
\begin{figure*}
\begin{center}
  \includegraphics[scale=0.8]{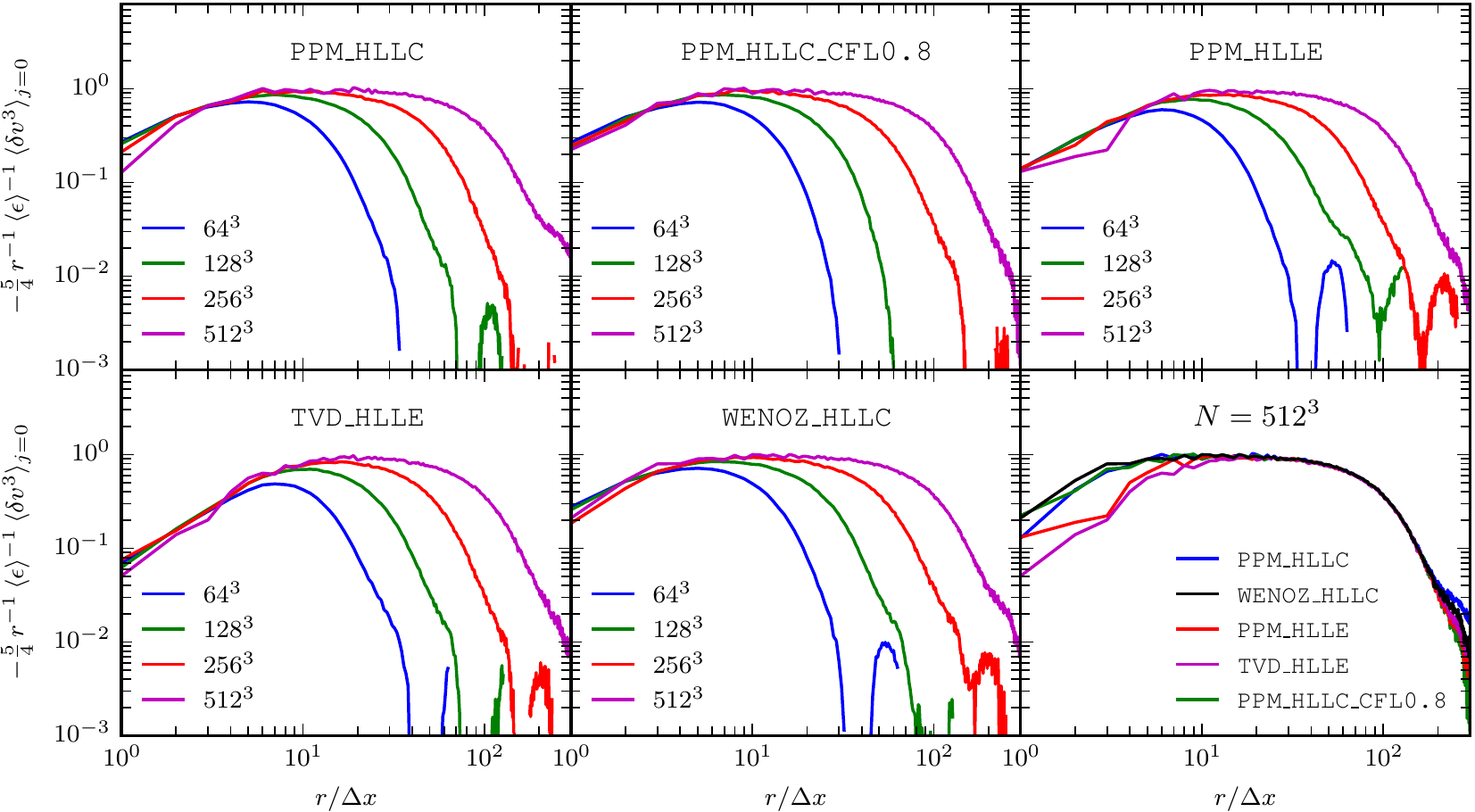}
  \caption{Compensated 3$^{\rm rd}$ order structure functions (equation
  \ref{eq:structure.functions}) for all the numerical methods and resolutions.
  The structure functions are compensated and scaled so that they should be
  close to one where the $4/5-$law (equation \ref{eq:45law}) is verified. The
  data is plotted as a function of the dimensionless separation $r/\Delta x$.
  The first five panels show the \texttt{PPM\_HLLC} (upper left),
  \texttt{PPM\_HLLC\_CFL0.8} (upper center), \texttt{PPM\_HLLE} (upper right),
  \texttt{TVD\_HLLE} (lower left) and \texttt{WENOZ\_HLLC} (lower center) group
  of runs. The last panel (lower right) shows a comparison of all of the methods
  at the highest resolution ($512^3$). The $4/5-$law is very well verified in
  our data suggesting that 1) anisotropic corrections are subdominant and 2) all
  of the simulations behave in a way consistent with large Reynolds numbers
  turbulence at the largest scales.
  \label{fig:structure.signed}}
\end{center}
\end{figure*}

The $4/5-$law (equation \ref{eq:45law}) is not a-priori valid in the regime of
turbulence we are considering. However, the $4/5-$law has been numerically
verified to hold also in some situations outside the domain of validity of its
derivation. For instance, for isotropic mildly compressible
decaying \cite{porter:02} and driven \cite{benzi:08} turbulence. In the
anisotropic case, however, anisotropic contributions cannot be
excluded \cite{biferale:02}, although they are known to be subdominant in some
important cases \cite{calzavarini:02, biferale:03, kaneda:08}. In this section we
show that equation \eqref{eq:45law} is consistent with our data over a wide
range of scales.

We compute the 3$^{\rm rd}$ order structure functions of the velocity, defined
by equation \eqref{eq:structure.functions}, in a rather simple way using
a random sample of 20,000 points in each of the $380$ 3D data dumps of our
simulations.  At each time, we compute the 3$^{\rm rd}$ power of the velocity
increments for each pair of points and accumulate and average in time the
results in bins of size $\Delta x$. The resulting structure functions are shown
in Figure \ref{fig:structure.signed}, compensated by $-\frac{5}{4} r^{-1}
\langle\epsilon\rangle^{-1}$, so that the resulting quantity should be equal to
one if the $4/5-$law is satisfied in our data. As was the case for the energy
spectra, we assume $\eta \sim \Delta x$ and plot the structure functions versus
$r/\Delta x$.

The degree with which the $4/5-$law is satisfied in our data is very good. We
see that anisotropic contributions only play a minor role in the
angle-integrated formulation of the $4/5-$law. This is in agreement with the
incompressible \ac{DNS} of  \cite{kaneda:08} and has been known to be true also
for Rayleigh-Bénard convection in most regimes \cite{lohse:10}. Our results
provide an important new example where this appears to hold true. Secondly, for
all of our simulations at $512^3$, we find
\begin{equation}
  \max_r \left\{ -\frac{5}{4} r^{-1} \langle\epsilon\rangle^{-1} \langle \delta v^3
  \rangle_{j=0} \right\}
\end{equation}
within $5\%$ of $1$. This level of accuracy is reached in \ac{DNS} simulations
achieving at least a Taylor micro-scale Reynolds number \cite{kaneda:08}
\begin{equation}
  R_\lambda = \frac{u' \lambda}{\nu} \sim 300\,,
\end{equation}
where $\nu$ is the kinematic viscosity, $u' = \frac{1}{\sqrt{3}}
v_{\mathrm{rms}}$ and $\lambda = (15 \nu u'^2/\langle\epsilon\rangle)^{1/2}$ is
the Taylor micro-scale. This corresponds to a large-scale Reynold numbers $R =
\frac{u' L}{\nu} \sim R_\lambda^2$, $L=1$ being the domain size, in excess of
$\sim$85,000. Reaching these Reynolds numbers in a \ac{DNS} requires resolutions
between $512^3$ and $1024^3$ using pseudo-spectral methods \cite{donzis:08}. This
large-scale estimate of the Reynolds number is consistent with previous
findings \cite{zhou:14}, although it is several orders of magnitude larger than
the one that could be naively estimated using $\nu_{\max}$. For instance, for
\texttt{PPM\_HLLC} at $512^3$, $\nu_{\max} \simeq 1.5 \times 10^{-3}$ and
$v_{\mathrm{rms}} \simeq 0.4$ giving
\begin{equation}
  \frac{u' L}{\nu_{\max}} \simeq 150\,.
\end{equation}
This apparent discrepancy is due to the fact that an \ac{ILES} is not a
\ac{DNS}. As a consequence, different quantities that in a \ac{DNS} depend on
the Reynolds number, such as the dissipation rate or the Kolmogorov scale,
behave as though the simulation had multiple values of the Reynolds number.

\section{The transverse spectrum}
\label{sec:results.transverse}
\begin{figure}
\begin{center}
  \includegraphics[scale=0.8]{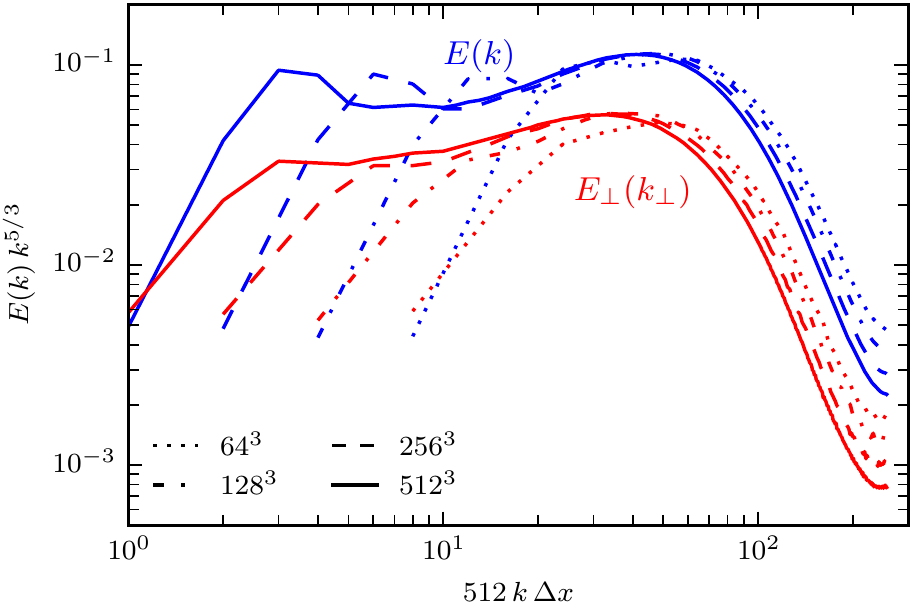}
  \caption{3D (equation \ref{eq:energy.spectrum}, blue) and 2D, transverse
  (equation \ref{eq:energy.spectrum.2d}, red) energy spectra for the
  \texttt{PPM\_HLLC} simulations. The energy spectra are compensated by
  $k^{5/3}$ to highlight eventual regions with Kolmogorov scaling. The spectra
  are plotted as a function of the dimensionless wave number $512\, k\, \Delta
  x$, as in Figure \ref{fig:energy.spectrum}. Although $E(k)$ and
  $E_\perp(k_\perp)$ have similar trends, the use of the transverse spectrum can
  overestimate the width of the bottleneck region.\label{fig:spectra.2d.vs.3d}}
\end{center}
\end{figure}

Finally, we want to comment on the use of 2D transverse spectra in 3D
simulations, a practice typically employed in the analysis of turbulence in
\ac{CCSN} simulations \cite{dolence:13, couch:14a, handy:14, abdikamalov:14b}.

Figure \ref{fig:spectra.2d.vs.3d} shows a comparison of the 2D transverse
spectrum $E_\perp(k_\perp)$ from equation \eqref{eq:energy.spectrum.2d} and the
3D energy spectrum from equation \eqref{eq:energy.spectrum} for the
\texttt{PPM\_HLLC} simulations.  The other runs (not shown) have the same
qualitative behavior.  We can see that the transverse spectrum follows
qualitatively the same trend as the 3D spectrum in terms of convergence. They
are both roughly compatible with a Kolmogorov scaling, but the bottleneck
appears to be more pronounced in the 2D spectrum than in the 3D spectrum. In
particular, $E_\perp(k_\perp)$ only shows a very small region that suggests an
inertial range, $3 \lesssim k \lesssim 5$ (as opposed to $5 \lesssim k \lesssim
10$ in $E(k)$).

 \cite{abdikamalov:14b} concluded, also based on the analysis of 2D spectra,
that turbulence in \ac{CCSN} simulations is dominated by the bottleneck effect.
Given the resolutions used in \ac{CCSN} studies, our work supports their
conclusion.  However, in the light of Figure \ref{fig:spectra.2d.vs.3d}, we
recommend that future studies supplement the analysis of 2D spectra with 3$^{\rm
rd}$ order structure functions, that, as we have shown, can give a more accurate
description of the energy cascade.

\section{Conclusions}
\label{sec:conclusions}
The details of the explosion mechanism of \ac{CCSN}e have eluded our
comprehension in spite of more than 50 years of studies \cite{woosley:05nature,
janka:12b, burrows:13a, foglizzo:15}. Recent numerical advances \cite{murphy:13,
couch:13d, couch:15a, mueller:15} suggest that turbulence might play a
fundamental role in tipping over the balance of the forces and lead to
successful explosions (see also Appendix \ref{sec:turbulent.pressure}). At the
same time, the level of accuracy of current simulations, which employ the
\ac{ILES} methodology, is unclear \cite{abdikamalov:14b}. Turbulence in
\ac{CCSN}e is mildly compressible, but strongly anisotropic \cite{murphy:13,
couch:15a}.  Simulations use rather dissipative numerical schemes (because they
have to deal with strong shock waves and complex microphysics) and relatively
low resolution, a combination (anisotropic turbulence and dissipative schemes)
that has not been systematically studied before.

With the goal of assessing the reliability of \ac{ILES} employed in
the study of \ac{CCSN}e, as well as in other areas of physics and
astrophysics, we performed a series of local simulations
of driven, anisotropic, weakly compressible turbulence. We compared five
commonly employed numerical schemes with different reconstruction
methods, Riemann solvers, and time step size. Each was run at $4$
different resolutions ranging from $64^3$ to $512^3$. Our analysis
focused on the fidelity with which the turbulent cascade is
represented in each model.  In particular, we performed an analysis
both in Fourier space (with the velocity power-spectra and the energy
flux) and in physical space (with the 3$^{\rm rd}$ order structure
functions). Finally, we measured the numerical viscosity of each
scheme from the residual of the specific kinetic energy equation.

We found that, on the one hand, all of the numerical setups are able to
accurately capture the decay rate of kinetic energy from the injection scale,
with errors at the few $\%$ level already at $128^3$ (\eg $\sim 2.5\%$ for
\texttt{PPM\_HLLC\_N128}). On the other hand, a large fraction of the energy is
at unresolved scales where it piles up due to the bottleneck effect and an
inertial range appears only at the highest resolutions ($512^3$). Even at this
resolution, which would be difficult to achieve in global simulations, only
roughly $\sim 80\%$ (the exact number depends on the scheme, see Section
\ref{sec:results.energy.cumulative}) of the energy is resolved, the remaining
$\sim 20\%$ accumulates as excess energy at intermediate scales (the cumulative
energy excess at the grid scale alone is as large as $\sim 5\%$ of the total
energy).

Current \ac{CCSN} simulations have resolutions of at most of $30 - 70$ points
covering the gain region \cite{abdikamalov:14b} (the energy injection scale).
Based on our analysis we expect that at these resolutions even the energy decay
rate from the largest scales will not be completely converged, but will show
errors of up to tens of percent, depending on the numerical scheme (see Section
\ref{sec:results.energy.decay}). At smaller scales, the dynamics is going to be
completely dominated by the bottleneck effect. This is in agreement with the
findings of \cite{abdikamalov:14b}, based on the use of global simulations
reaching a maximum resolution of $66$ grid points covering radially the extent
of the gain region.

Based on our findings, we expect that, if the resolution in global simulations
is increased by a factor $\sim 2$ from the one of \cite{abdikamalov:14b}, the
decay rate will be converged to within a few $\%$ of the asymptotic value. This
implies that the ratio between thermal and kinetic energy, a crucial quantity
for the onset of the explosion, will also be converged to within a few $\%$, at
least when the energy injection rate changes slowly compared to the eddy
turnover time (which is roughly $\sim 20\,\mathrm{ms}$ in a
\ac{CCSN} \cite{ott:13a, couch:14a}).  Unfortunately, while the lead up to
explosion occurs over a larger timescale of a few hundred milliseconds, the
transition to explosion can happen over much shorter timescales (one turnover
time or less) \cite{couch:15a}.  This means that the dynamics of the cascade over
smaller time and length scales in the gain region also needs to be captured
correctly since changes in the energy input rate on such short time scales will
yield an inaccurate representation of the energy on large scales due to
the bottleneck effect. This could require an increase of resolution by a factor
$\sim 4 - 8$ with respect to current high-resolution simulations. Additional
work using semi-global or global simulations will be required to more firmly
establish the resolutions requirements at the transition of the explosion.

Concerning the properties of anisotropic turbulence in our
simulations, we found anisotropy contributions to the energy spectrum
and to the angle-averaged formulation of the $4/5-$law to be
subdominant: the accuracy with which the $4/5-$law is satisfied is
limited only by the employed resolution and the energy spectrum
appears to be consistent with Kolmogorov $k^{-5/3}$ scaling.  We also
found the transverse energy spectrum with respect to the direction of
anisotropy, a quantity typically computed in \ac{CCSN} simulations, to
overestimate the bottleneck with respect to the angle-integrated 3D
spectrum.  For this reason, we recommend future studies of \ac{CCSN}
to supplement (or replace) the analysis of the transverse spectrum
with the analysis of the 3$^{\rm rd}$ order, angle-integrated,
structure function (or, where possible, with the 3D spectrum itself).

Our results are, of course, dependent on the choice of the numerical scheme. In
particular, we found significant differences in the dissipative properties of
schemes employing the HLLC Riemann solver with respect to schemes using the more
dissipative HLLE solver. The reconstruction order of the scheme is also
important, although, while significant differences are found between TVD and
PPM, the differences between PPM and WENOZ are much more minute (despite WENOZ
being significantly more computationally expensive than PPM). In the end, none
of the schemes we considered seems to be able to yield an accurate
representation of the kinetic energy distribution across different scales at an
affordable resolution for global \ac{CCSN} simulations. A possible way forward
would be to adopt low-dissipation numerical schemes especially designed for the
use in \ac{ILES}, such as the methods proposed by \cite{hickel:06, martin:06}
or \cite{thornber:08}. Implementing and testing these schemes will be subject
of future work.

An important limitation of the present work is that we considered a
very idealized setup. On the one hand, this allows us to benchmark the
behavior of \ac{ILES} in a controlled environment. On the other hand,
our simulations cannot fully capture all features of the turbulent
convective flow in a \ac{CCSN}.  Unlike the situation in a CCSN, our
local simulations did not include a vertical advective velocity field
that is due to the accretion of the stellar mantle.  However, the
advective velocities are nearly constant in the regions of interest
and Galilean invariance ensures that our results are unaffected. More
limiting is the local nature of our simulations and the inevitable
choice of boundary conditions. Moreover, our simulations could not
take into account spatial variations in gravity and the large-scale
radial convergence of the flow in globally spherical problems like
collapsing stars. Addressing these issues will also be subject of
future work.

\appendix

\section{The role of turbulence in core-collapse supernova explosions}
\label{sec:turbulent.pressure}
In this appendix we present a brief discussion of the importance of
turbulent pressure in the explosion of massive stars. To set the
stage, we will briefly summarize what is known of the dynamics of the
most common class of \ac{CCSN}e that are relevant for our later
discussion. This is done for the benefit of readers that are not
supernova specialists and it is not meant to be a complete review of
the status of the field, for which we refer, instead, to the reviews
of \cite{janka:12b} and \cite{burrows:13a}.
Next, we will discuss the role of turbulence and, in particular, of
turbulent pressure on the explosion mechanism, in light of some recent
results \cite{murphy:13, couch:15a}.

\subsection{The neutrino mechanism}
Towards the end of their evolution, massive stars form massive ($\sim
1.5$ solar mass) iron cores at their center. Since the iron nucleus
has the largest binding energy per nucleon, no energy can be extracted
from nuclear fusion beyond iron. The iron core is essentially inert
and supported against gravity only by the degeneracy pressure of
relativistic electrons. The mass of the iron core increases with time
as more iron-group material is added by silicon shell burning.
Electron capture on protons, which becomes energetically favorable at
high densities, depletes the core of electrons and thus reduces the
pressure supporting it against gravity. Eventually, the core becomes
dynamically unstable and collapses.

During the collapse, the subsonically collapsing inner core ($\sim
0.5$ solar masses) contracts until it reaches densities comparable to
that in atomic nuclei ($\sim 4-5 \times 10^{14}\,
\mathrm{g}\,\mathrm{cm}^{-3}$). At this point, the nuclear equation of
state stiffens (due to the strong nuclear force). This halts the
collapse of the inner core. It stops, bounces back and a \ac{PNS} is
formed.  The outer core, however, is still collapsing supersonically
and a strong shock wave is launched at the interface between the inner
and outer core.

It was once thought that this shock wave would travel outwards
dynamically, depositing its energy in the outer layers of the star,
causing the explosion.  However, multiple numerical simulations
performed over multiple decades have consistently shown that the
initial shock fails to explode the star. Instead, it stalls due to energy
losses to the dissociation of heavy nuclei into free nucleons and to
the emission of neutrinos that stream away from the
neutrino-semitransparent regions behind the shock \cite{bethe:90}. The
shock generally stalls within only a few tens of milliseconds of core
bounce and turns into an accretion shock standing at a radius of $\sim
100-200\ \mathrm{km}$. The accretion rate through the shock is so high
(a fraction of a solar mass per second) that, if nothing revitalizes
the shock within $\sim 1-2$ seconds, the gravitational force would
overwhelm the nuclear repulsion force, collapsing the core of the
supernova to a black hole, precluding explosion (e.g., \cite{oconnor:11}).

During this time, however, the \ac{PNS} will release a significant
fraction of its binding energy in the form of neutrinos (of order
$10^{46}\, \mathrm{J}$). Converting a few percent of that energy into
kinetic energy would be enough to unbind the stellar envelope and
power the supernova explosion. In the standard neutrino mechanism it
is theorized that a small fraction neutrinos emitted from the edge of
the protoneutron star is re-absorbed in the region right behind the
stalled shock. The deposition of neutrino energy leads to higher
thermal pressure so that the shock can eventually overcome the ram
pressure of accretion and accelerates in a run-away
process \cite{bethe:90,pejcha:12a}. Turbulence in the heating region
behind the shock increases the time a fluid parcel spends in that
region and, importantly, turbulent pressure helps in overcoming the
ram pressure of accretion (see next Section and \cite{couch:14a}). It
is, however, presently unclear if neutrino heating (even if aided by
turbulence in launching the explosion) is able to provide enough
energy to power the explosions to the energies inferred from
astronomical observations.

\subsection{Turbulent pressure and the Rankine-Hugoniot conditions}
Simulations \cite{bhf:95,murphy:13, couch:15a} have shown that turbulence and,
in particular, turbulent pressure behind the shock, could play an important role
in aiding the explosion.  To see why this is the case, we consider the
Rankine-Hugoniot momentum condition for a standing accretion shock in a
supernova core,
\begin{equation}\label{eq:rankine.hugoniot}
  s [ \rho_d v_d^r - \rho_u v_u^r ] =
  \rho_d (v_d^r)^2 + p_d - \rho_u (v_u^r)^2 - p_u,
\end{equation}
where $s$ is the shock speed and $\cdot_d$ and $\cdot_u$ denote the downstream
and upstream values respectively. For the purpose of our discussion, we can
assume the upstream gas to be cold and free-falling:
\begin{align}
  p_u = 0 && (v_u^r)^2 = \frac{GM}{r}.
\end{align}
For the shock to expand we must then have
\begin{equation}\label{eq:rankine.hugoniot.simple}
  \rho_d (v_d^r)^2 + p_d > \rho_u \frac{GM}{r}.
\end{equation}
In the presence of turbulence, \cite{murphy:13} suggested to modify equation
\eqref{eq:rankine.hugoniot.simple} in a way akin to a Reynolds decomposition and
write it as
\begin{equation}\label{eq:rankine.hugoniot.turb}
  \rho_d (\bar{v}_d^r)^2 + \rho_d (\delta v_d^r)^2 + p_d > \rho_u \frac{GM}{r},
\end{equation}
where $\mathbf{v}$ is the average velocity and $\delta\mathbf{v} =
\bar{\mathbf{v}} - \mathbf{v}$ is the turbulent velocity. Although not entirely
rigorous, equation \eqref{eq:rankine.hugoniot.turb} has been shown to be well
verified in the numerical simulations if angular averages are used to compute
the respective quantities \cite{murphy:13, couch:15a}.
\cite{couch:15a} have shown that the turbulent pressure expressed in
this fashion can exceed 50\% of the thermal pressure, making a
very significant contribution to the momentum balance in \eqref{eq:rankine.hugoniot.turb}.

Going beyond the arguments of \cite{murphy:13}, we can reinterpret equation
\eqref{eq:rankine.hugoniot.turb} as being the Rankine-Hugoniot condition for a
fluid with a modified equation of state, which has two separate internal degrees
of freedom: thermodynamical and turbulent. To this aim, we express $\delta
v_d^r$ in terms of the specific turbulent energy 
\begin{equation}
  e_{\mathrm{turb}} =  \frac{1}{2} | \delta\mathbf{v} |^2
\end{equation}
noting that
\begin{equation}
  |\delta\mathbf{v}|^2 := (\delta v^r)^2 + (\delta v^\theta)^2 +
  (\delta v^\phi)^2,
\end{equation}
and using the fact that
\begin{equation}\label{eq:turbulence.anisotropy}
  (\delta v^r)^2 \simeq (\delta v^\theta)^2 + (\delta v^\phi)^2
\end{equation}
in \ac{CCSN} turbulence, to obtain
\begin{equation}\label{eq:vr.vs.turb.epsilon}
  (\delta v^r)^2 \simeq \frac{1}{2} | \delta\mathbf{v} |^2 =
  e_{\mathrm{turb}}.
\end{equation}
Assuming the pressure varies like $p = (\gamma - 1) \rho e$, and
substituting \eqref{eq:vr.vs.turb.epsilon} into
\eqref{eq:rankine.hugoniot.turb}, we find
\begin{equation}
  \rho_d (\bar{v}_d^r)^2 + (\gamma_{\mathrm{th}} - 1) \rho_d e_d +
  (\gamma_{\mathrm{turb}} - 1) \rho_d e_{\mathrm{turb}} > \rho_u
  \frac{GM}{r},
\end{equation}
where $\gamma_{\mathrm{th}} \simeq 4/3$ is the thermodynamical adiabatic index,
$e_d$ is the downstream thermal energy and $\gamma_{\mathrm{turb}} = 2$ is the
equivalent adiabatic index associated with anisotropic \ac{CCSN} turbulence.
Since $\gamma_{\mathrm{turb}} > \gamma_{\mathrm{th}}$, we see that turbulent
energy is more efficient, per unit specific internal energy, at pushing the
shock than thermal energy.

We point out that, if equation \eqref{eq:turbulence.anisotropy} is dropped and
turbulence is assumed to be isotropic, then $\gamma_{\mathrm{turb}} = 5/3$,
which is still larger than $\gamma_{\mathrm{th}}$, but not as large as for the
anisotropic case. This is a simple consequence of the fact that anisotropic
turbulence has an anisotropic pressure, which is stronger in the radial
direction.

In both cases, since the total energy is conserved, the relevant quantity is the
ratio $e_{\mathrm{turb}}/e$. From standard turbulent phenomenology we expect
that this ratio will only depend on macroscopic parameters, such as the net
heating rate, the accretion rate and so on, and not on the details of the
viscosity. For this reason, we expect this ratio to be correctly captured in
\ac{ILES} achieving a sufficiently high resolution.

As a final remark, we point out that a similar argument has been
recently proposed by \cite{mueller:15} who formulated their equations
in terms of the turbulent Mach number, as opposed to the turbulent
energy.

\begin{backmatter}

\section*{Competing interests}
The authors declare that they have no competing interests.

\section*{Author's contributions}
DR ran the simulations, performed the analysis of the data and wrote the basic
draft of this paper. SMC assisted with the use of the FLASH code. CDO had the
original idea that started this investigation.  All of the authors took part in
discussions concerning the results and contributed corrections and improvements
on the early draft of the manuscript.

\section*{Acknowledgements}
We acknowledge helpful discussions with E.~Abdikamalov, W.~D.~Arnett,
A.~Burrows, R.~Fisher, C.~Meakin, P.~M\"osta, J.~Murphy, M.~Norman, and
L.~Roberts. This research was partially supported by the National Science
Foundation under award nos.\ AST-1212170 and PHY-1151197 and by the Sherman
Fairchild Foundation. The simulations were performed on the Caltech compute
cluster Zwicky (NSF MRI-R2 award no.\ PHY-0960291), on the NSF XSEDE network
under allocation TG-PHY100033, and on NSF/NCSA BlueWaters under NSF PRAC award
no.\ ACI-1440083. The software used in this work was in part developed by the
DOE NNSA-ASC OASCR Flash Center at the University of Chicago.


\newcommand{\BMCxmlcomment}[1]{}

\BMCxmlcomment{

<refgrp>

<bibl id="B1">
  <title><p>{The Black Hole Formation Probability}</p></title>
  <aug>
    <au><snm>{Clausen}</snm><fnm>D.</fnm></au>
    <au><snm>{Piro}</snm><fnm>A. L.</fnm></au>
    <au><snm>{Ott}</snm><fnm>C. D.</fnm></au>
  </aug>
  <source>\apj</source>
  <pubdate>2015</pubdate>
  <volume>799</volume>
  <fpage>190</fpage>
</bibl>

<bibl id="B2">
  <title><p>{The physics of core-collapse supernovae}</p></title>
  <aug>
    <au><snm>{Woosley}</snm><fnm>S. E.</fnm></au>
    <au><snm>{Janka}</snm><fnm>H. T.</fnm></au>
  </aug>
  <source>Nature Physics</source>
  <pubdate>2005</pubdate>
  <volume>1</volume>
  <fpage>147</fpage>
</bibl>

<bibl id="B3">
  <title><p>{Core-collapse supernovae: Reflections and directions}</p></title>
  <aug>
    <au><snm>{Janka}</snm><fnm>H. T.</fnm></au>
    <au><snm>{Hanke}</snm><fnm>F.</fnm></au>
    <au><snm>{H{\"u}depohl}</snm><fnm>L.</fnm></au>
    <au><snm>{Marek}</snm><fnm>A.</fnm></au>
    <au><snm>{M{\"u}ller}</snm><fnm>B.</fnm></au>
    <au><snm>{Obergaulinger}</snm><fnm>M.</fnm></au>
  </aug>
  <source>Prog. Th. Exp. Phys.</source>
  <pubdate>2012</pubdate>
  <volume>2012</volume>
  <issue>1</issue>
  <fpage>010000</fpage>
</bibl>

<bibl id="B4">
  <title><p>{Colloquium: Perspectives on core-collapse supernova
  theory}</p></title>
  <aug>
    <au><snm>{Burrows}</snm><fnm>A.</fnm></au>
  </aug>
  <source>Rev. Mod. Phys.</source>
  <pubdate>2013</pubdate>
  <volume>85</volume>
  <fpage>245</fpage>
</bibl>

<bibl id="B5">
  <title><p>{The explosion mechanism of core-collapse supernovae: progress in
  supernova theory and experiments}</p></title>
  <aug>
    <au><snm>Foglizzo</snm><fnm>T</fnm></au>
    <au><snm>Kazeroni</snm><fnm>R</fnm></au>
    <au><snm>Guilet</snm><fnm>J</fnm></au>
    <au><snm>Masset</snm><fnm>F</fnm></au>
    <au><snm>Gonz\'{a}lez</snm><fnm>M</fnm></au>
    <au><cnm>Others</cnm></au>
  </aug>
  <pubdate>2015</pubdate>
</bibl>

<bibl id="B6">
  <title><p>{Explosion Mechanisms of Core-Collapse Supernovae}</p></title>
  <aug>
    <au><snm>{Janka}</snm><fnm>H. T.</fnm></au>
  </aug>
  <source>Ann. Rev. Nuc. Par. Sci.</source>
  <pubdate>2012</pubdate>
  <volume>62</volume>
  <fpage>407</fpage>
</bibl>

<bibl id="B7">
  <title><p>{The convective engine paradigm for the supernova explosion
  mechanism and its consequences.}</p></title>
  <aug>
    <au><snm>{Herant}</snm><fnm>M.</fnm></au>
  </aug>
  <source>Phys.~Rep.</source>
  <pubdate>1995</pubdate>
  <volume>256</volume>
  <fpage>117</fpage>
</bibl>

<bibl id="B8">
  <title><p>{On the Nature of Core-Collapse Supernova Explosions}</p></title>
  <aug>
    <au><snm>{Burrows}</snm><fnm>A.</fnm></au>
    <au><snm>{Hayes}</snm><fnm>J.</fnm></au>
    <au><snm>{Fryxell}</snm><fnm>B. A.</fnm></au>
  </aug>
  <source>\apj</source>
  <pubdate>1995</pubdate>
  <volume>450</volume>
  <fpage>830</fpage>
</bibl>

<bibl id="B9">
  <title><p>{Neutrino heating, convection, and the mechanism of Type-II
  supernova explosions.}</p></title>
  <aug>
    <au><snm>{Janka}</snm><fnm>H. T.</fnm></au>
    <au><snm>{M\"uller}</snm><fnm>E.</fnm></au>
  </aug>
  <source>\aap</source>
  <pubdate>1996</pubdate>
  <volume>306</volume>
  <fpage>167</fpage>
</bibl>

<bibl id="B10">
  <title><p>{Neutrino-driven Convection versus Advection in Core-Collapse
  Supernovae}</p></title>
  <aug>
    <au><snm>{Foglizzo}</snm><fnm>T.</fnm></au>
    <au><snm>{Scheck}</snm><fnm>L.</fnm></au>
    <au><snm>{Janka}</snm><fnm>H. T.</fnm></au>
  </aug>
  <source>\apj</source>
  <pubdate>2006</pubdate>
  <volume>652</volume>
  <fpage>1436</fpage>
</bibl>

<bibl id="B11">
  <title><p>{Stability of Standing Accretion Shocks, with an Eye toward
  Core-Collapse Supernovae}</p></title>
  <aug>
    <au><snm>{Blondin}</snm><fnm>J. M.</fnm></au>
    <au><snm>{Mezzacappa}</snm><fnm>A.</fnm></au>
    <au><snm>{DeMarino}</snm><fnm>C.</fnm></au>
  </aug>
  <source>\apj</source>
  <pubdate>2003</pubdate>
  <volume>584</volume>
  <fpage>971</fpage>
</bibl>

<bibl id="B12">
  <title><p>{Instability of a Stalled Accretion Shock: Evidence for the
  Advective-Acoustic Cycle}</p></title>
  <aug>
    <au><snm>{Foglizzo}</snm><fnm>T.</fnm></au>
    <au><snm>{Galletti}</snm><fnm>P.</fnm></au>
    <au><snm>{Scheck}</snm><fnm>L.</fnm></au>
    <au><snm>{Janka}</snm><fnm>H. T.</fnm></au>
  </aug>
  <source>\apj</source>
  <pubdate>2007</pubdate>
  <volume>654</volume>
  <fpage>1006</fpage>
</bibl>

<bibl id="B13">
  <title><p>{Neutrino-driven Turbulent Convection and Standing Accretion Shock
  Instability in Three-dimensional Core-collapse Supernovae}</p></title>
  <aug>
    <au><snm>{Abdikamalov}</snm><fnm>E.</fnm></au>
    <au><snm>{Ott}</snm><fnm>C. D.</fnm></au>
    <au><snm>{Radice}</snm><fnm>D.</fnm></au>
    <au><snm>{Roberts}</snm><fnm>L. F.</fnm></au>
    <au><snm>{Haas}</snm><fnm>R.</fnm></au>
    <au><snm>{Reisswig}</snm><fnm>C.</fnm></au>
    <au><snm>{M{\"o}sta}</snm><fnm>P.</fnm></au>
    <au><snm>{Klion}</snm><fnm>H.</fnm></au>
    <au><snm>{Schnetter}</snm><fnm>E.</fnm></au>
  </aug>
  <source>\apj</source>
  <pubdate>2015</pubdate>
  <volume>808</volume>
  <fpage>70</fpage>
</bibl>

<bibl id="B14">
  <title><p>{The Dominance of Neutrino-driven Convection in Core-collapse
  Supernovae}</p></title>
  <aug>
    <au><snm>{Murphy}</snm><fnm>J. W.</fnm></au>
    <au><snm>{Dolence}</snm><fnm>J. C.</fnm></au>
    <au><snm>{Burrows}</snm><fnm>A.</fnm></au>
  </aug>
  <source>\apj</source>
  <pubdate>2013</pubdate>
  <volume>771</volume>
  <fpage>52</fpage>
</bibl>

<bibl id="B15">
  <title><p>{The Role of Turbulence in Neutrino-driven Core-collapse Supernova
  Explosions}</p></title>
  <aug>
    <au><snm>{Couch}</snm><fnm>S. M.</fnm></au>
    <au><snm>{Ott}</snm><fnm>C. D.</fnm></au>
  </aug>
  <source>\apj</source>
  <pubdate>2015</pubdate>
  <volume>799</volume>
  <fpage>5</fpage>
</bibl>

<bibl id="B16">
  <title><p>{Large Eddy Simulation for Compressible Flows}</p></title>
  <aug>
    <au><snm>{Garnier}</snm><fnm>E.</fnm></au>
    <au><snm>{Adams}</snm><fnm>N.</fnm></au>
    <au><snm>{Sagaut}</snm><fnm>P.</fnm></au>
  </aug>
  <source>Large Eddy Simulation for Compressible Flows, by Garnier et al., S.
  B.~ISBN 978-90-481-2819-8.~Published by Springer, 2009.</source>
  <publisher>Springer, Berlin, Germany</publisher>
  <pubdate>2000</pubdate>
</bibl>

<bibl id="B17">
  <title><p>{Implicit Large Eddy Simulation: Computing Turbulent Fluid
  Dynamics}</p></title>
  <aug>
    <au><snm>Grinstein</snm><fnm>FF</fnm></au>
    <au><snm>Margolin</snm><fnm>LG</fnm></au>
    <au><snm>Rider</snm><fnm>WJ</fnm></au>
  </aug>
  <publisher>Cambridge University Press</publisher>
  <pubdate>2011</pubdate>
  <fpage>578</fpage>
</bibl>

<bibl id="B18">
  <title><p>Hidden conservation laws in hydrodynamics; energy and dissipation
  rate fluctuation spectra in strong turbulence</p></title>
  <aug>
    <au><snm>Yakhot</snm><fnm>V</fnm></au>
    <au><snm>Zakharov</snm><fnm>V</fnm></au>
  </aug>
  <source>Phys. D</source>
  <pubdate>1993</pubdate>
  <volume>64</volume>
  <issue>4</issue>
  <fpage>379</fpage>
</bibl>

<bibl id="B19">
  <title><p>On the universal form of energy spectra in fully developed
  turbulence</p></title>
  <aug>
    <au><snm>She</snm><fnm>Z</fnm></au>
    <au><snm>Jackson</snm><fnm>E</fnm></au>
  </aug>
  <source>Phys. Fluids</source>
  <pubdate>1993</pubdate>
  <volume>5</volume>
  <issue>7</issue>
  <fpage>1526</fpage>
</bibl>

<bibl id="B20">
  <title><p>Bottleneck phenomenon in developed turbulence</p></title>
  <aug>
    <au><snm>Falkovich</snm><fnm>G</fnm></au>
  </aug>
  <source>Phys. Fluids</source>
  <pubdate>1994</pubdate>
  <volume>6</volume>
  <issue>4</issue>
  <fpage>1411</fpage>
  <lpage>1414</lpage>
</bibl>

<bibl id="B21">
  <title><p>Energy transfer and bottleneck effect in turbulence</p></title>
  <aug>
    <au><snm>Verma</snm><fnm>MK</fnm></au>
    <au><snm>Donzis</snm><fnm>D</fnm></au>
  </aug>
  <source>J. Phys. A</source>
  <pubdate>2007</pubdate>
  <volume>40</volume>
  <issue>16</issue>
  <fpage>4401</fpage>
</bibl>

<bibl id="B22">
  <title><p>{Hyperviscosity, Galerkin Truncation, and Bottlenecks in
  Turbulence}</p></title>
  <aug>
    <au><snm>Frisch</snm><fnm>U</fnm></au>
    <au><snm>Kurien</snm><fnm>S</fnm></au>
    <au><snm>Pandit</snm><fnm>R</fnm></au>
    <au><snm>Pauls</snm><fnm>W</fnm></au>
    <au><snm>Ray</snm><fnm>S</fnm></au>
    <au><snm>Wirth</snm><fnm>A</fnm></au>
    <au><snm>Zhu</snm><fnm>JZ</fnm></au>
  </aug>
  <source>\prl</source>
  <pubdate>2008</pubdate>
  <volume>101</volume>
  <issue>14</issue>
  <fpage>144501</fpage>
  <url>http://link.aps.org/doi/10.1103/PhysRevLett.101.144501</url>
</bibl>

<bibl id="B23">
  <title><p>{High-resolution Three-dimensional Simulations of Core-collapse
  Supernovae in Multiple Progenitors}</p></title>
  <aug>
    <au><snm>{Couch}</snm><fnm>S. M.</fnm></au>
    <au><snm>{O'Connor}</snm><fnm>E. P.</fnm></au>
  </aug>
  <source>\apj</source>
  <pubdate>2014</pubdate>
  <volume>785</volume>
  <fpage>123</fpage>
</bibl>

<bibl id="B24">
  <title><p>{Convergence Tests for the Piecewise Parabolic Method and
  Navier-Stokes Solutions for Homogeneous Compressible Turbulence}</p></title>
  <aug>
    <au><snm>{Sytine}</snm><fnm>I. V.</fnm></au>
    <au><snm>{Porter}</snm><fnm>D. H.</fnm></au>
    <au><snm>{Woodward}</snm><fnm>P. R.</fnm></au>
    <au><snm>{Hodson}</snm><fnm>S. W.</fnm></au>
    <au><snm>{Winkler}</snm><fnm>K. H.</fnm></au>
  </aug>
  <source>J. Comp. Phys.</source>
  <pubdate>2000</pubdate>
  <volume>158</volume>
  <fpage>225</fpage>
</bibl>

<bibl id="B25">
  <title><p>{On the use of shock-capturing schemes for large-eddy
  simulation}</p></title>
  <aug>
    <au><snm>Garnier</snm><fnm>E</fnm></au>
    <au><snm>Mossi</snm><fnm>M</fnm></au>
    <au><snm>Sagaut</snm><fnm>P</fnm></au>
  </aug>
  <source>Journal of Computational \ldots</source>
  <pubdate>1999</pubdate>
  <volume>311</volume>
  <fpage>273</fpage>
  <lpage>-311</lpage>
  <url>http://www.sciencedirect.com/science/article/pii/S002199919996268X</url>
</bibl>

<bibl id="B26">
  <title><p>{Assessment of high-resolution methods for numerical simulations of
  compressible turbulence with shock waves}</p></title>
  <aug>
    <au><snm>Johnsen</snm><fnm>E</fnm></au>
    <au><snm>Larsson</snm><fnm>J</fnm></au>
    <au><snm>Bhagatwala</snm><fnm>AV</fnm></au>
    <au><snm>Cabot</snm><fnm>WH</fnm></au>
    <au><snm>Moin</snm><fnm>P</fnm></au>
    <au><snm>Olson</snm><fnm>BJ</fnm></au>
    <au><snm>Rawat</snm><fnm>PS</fnm></au>
    <au><snm>Shankar</snm><fnm>SK</fnm></au>
    <au><snm>Sj\"{o}green</snm><fnm>B</fnm></au>
    <au><snm>Yee</snm><fnm>H.C.</fnm></au>
    <au><snm>Zhong</snm><fnm>X</fnm></au>
    <au><snm>Lele</snm><fnm>SK</fnm></au>
  </aug>
  <source>Journal of Computational Physics</source>
  <publisher>Elsevier Inc.</publisher>
  <pubdate>2010</pubdate>
  <volume>229</volume>
  <issue>4</issue>
  <fpage>1213</fpage>
  <lpage>-1237</lpage>
  <url>http://linkinghub.elsevier.com/retrieve/pii/S0021999109005804</url>
</bibl>

<bibl id="B27">
  <title><p>{Turbulent Convection in Stellar Interiors. II. The Velocity
  Field}</p></title>
  <aug>
    <au><snm>{Arnett}</snm><fnm>D.</fnm></au>
    <au><snm>{Meakin}</snm><fnm>C.</fnm></au>
    <au><snm>{Young}</snm><fnm>P. A.</fnm></au>
  </aug>
  <source>\apj</source>
  <pubdate>2009</pubdate>
  <volume>690</volume>
  <fpage>1715</fpage>
</bibl>

<bibl id="B28">
  <title><p>{The residual anisotropy at small scales in high shear
  turbulence}</p></title>
  <aug>
    <au><snm>Casciola</snm><fnm>C. M.</fnm></au>
    <au><snm>Gualtieri</snm><fnm>P.</fnm></au>
    <au><snm>Jacob</snm><fnm>B.</fnm></au>
    <au><snm>Piva</snm><fnm>R.</fnm></au>
  </aug>
  <source>Physics of Fluids</source>
  <pubdate>2007</pubdate>
  <volume>19</volume>
  <issue>10</issue>
  <fpage>101704</fpage>
  <url>http://scitation.aip.org/content/aip/journal/pof2/19/10/10.1063/1.2800043</url>
</bibl>

<bibl id="B29">
  <title><p>{The Piecewise Parabolic Method (PPM) for Gas-Dynamical
  Simulations}</p></title>
  <aug>
    <au><snm>{Colella}</snm><fnm>P.</fnm></au>
    <au><snm>{Woodward}</snm><fnm>P. R.</fnm></au>
  </aug>
  <source>J. Comp. Phys.</source>
  <pubdate>1984</pubdate>
  <volume>54</volume>
  <fpage>174</fpage>
</bibl>

<bibl id="B30">
  <title><p>{R}iemann {S}olvers and {N}umerical {M}ethods for {F}luid
  {D}ynamics</p></title>
  <aug>
    <au><snm>Toro</snm><fnm>E. F.</fnm></au>
  </aug>
  <publisher>Berlin: Springer</publisher>
  <pubdate>1999</pubdate>
</bibl>

<bibl id="B31">
  <title><p>Numerical dissipation and the bottleneck effect in simulations of
  compressible isotropic turbulence</p></title>
  <aug>
    <au><snm>Schmidt</snm><fnm>W</fnm></au>
    <au><snm>Hillebrandt</snm><fnm>W</fnm></au>
    <au><snm>Niemeyer</snm><fnm>JC</fnm></au>
  </aug>
  <source>Computers \& Fluids</source>
  <pubdate>2006</pubdate>
  <volume>35</volume>
  <issue>4</issue>
  <fpage>353</fpage>
</bibl>

<bibl id="B32">
  <title><p>{On the implicit large eddy simulations of homogeneous decaying
  turbulence}</p></title>
  <aug>
    <au><snm>Thornber</snm><fnm>B</fnm></au>
    <au><snm>Mosedale</snm><fnm>A</fnm></au>
    <au><snm>Drikakis</snm><fnm>D</fnm></au>
  </aug>
  <source>Journal of Computational Physics</source>
  <pubdate>2007</pubdate>
  <volume>226</volume>
  <issue>2</issue>
  <fpage>1902</fpage>
  <lpage>-1929</lpage>
  <url>http://linkinghub.elsevier.com/retrieve/pii/S0021999107002690</url>
</bibl>

<bibl id="B33">
  <title><p>{FLASH: An Adaptive Mesh Hydrodynamics Code for Modeling
  Astrophysical Thermonuclear Flashes}</p></title>
  <aug>
    <au><snm>{Fryxell}</snm><fnm>B.</fnm></au>
    <au><snm>{Olson}</snm><fnm>K.</fnm></au>
    <au><snm>{Ricker}</snm><fnm>P.</fnm></au>
    <au><snm>{Timmes}</snm><fnm>F. X.</fnm></au>
    <au><snm>{Zingale}</snm><fnm>M.</fnm></au>
    <au><snm>{Lamb}</snm><fnm>D. Q.</fnm></au>
    <au><snm>{MacNeice}</snm><fnm>P.</fnm></au>
    <au><snm>{Rosner}</snm><fnm>R.</fnm></au>
    <au><snm>{Truran}</snm><fnm>J. W.</fnm></au>
    <au><snm>{Tufo}</snm><fnm>H.</fnm></au>
  </aug>
  <source>\apjs</source>
  <pubdate>2000</pubdate>
  <volume>131</volume>
  <fpage>273</fpage>
</bibl>

<bibl id="B34">
  <title><p>Extensible component-based architecture for FLASH, a massively
  parallel, multiphysics simulation code</p></title>
  <aug>
    <au><snm>Dubey</snm><fnm>A.</fnm></au>
    <au><snm>Antypas</snm><fnm>K.</fnm></au>
    <au><snm>Ganapathy</snm><fnm>M. K.</fnm></au>
    <au><snm>Reid</snm><fnm>L. B.</fnm></au>
    <au><snm>Riley</snm><fnm>K.</fnm></au>
    <au><snm>Sheeler</snm><fnm>D.</fnm></au>
    <au><snm>Siegel</snm><fnm>A.</fnm></au>
    <au><snm>Weide</snm><fnm>K.</fnm></au>
  </aug>
  <source>Parallel Comput.</source>
  <pubdate>2009</pubdate>
  <volume>35</volume>
  <fpage>512</fpage>
</bibl>

<bibl id="B35">
  <title><p>FLASH: A Multi-physics Code for Adaptive Mesh Computational Fluid
  Dynamics in Astrophysics</p></title>
  <aug>
    <au><snm>{Lee}</snm><fnm>D.</fnm></au>
    <au><snm>{Tzeferacos}</snm><fnm>P.</fnm></au>
    <au><snm>{Couch}</snm><fnm>S. M.</fnm></au>
    <au><snm>{Bachan}</snm><fnm>J.</fnm></au>
    <au><snm>{Daley}</snm><fnm>C.</fnm></au>
    <au><snm>{Fatenejad}</snm><fnm>M.</fnm></au>
    <au><snm>{Flocke}</snm><fnm>N.</fnm></au>
    <au><snm>{Lamb}</snm><fnm>D.</fnm></au>
    <au><snm>{Weide}</snm><fnm>K.</fnm></au>
    <au><snm>{Dubey}</snm><fnm>A.</fnm></au>
  </aug>
  <source>To be submitted to \apjs</source>
  <pubdate>2014</pubdate>
</bibl>

<bibl id="B36">
  <title><p>{Multidimensional upwind methods for hyperbolic conservation
  laws}</p></title>
  <aug>
    <au><snm>{Colella}</snm><fnm>P.</fnm></au>
  </aug>
  <source>J.\ Comput.\ Phys.</source>
  <pubdate>1990</pubdate>
  <volume>87</volume>
  <fpage>171</fpage>
  <lpage>200</lpage>
</bibl>

<bibl id="B37">
  <title><p>{An unsplit staggered mesh scheme for multidimensional
  magnetohydrodynamics}</p></title>
  <aug>
    <au><snm>{Lee}</snm><fnm>D.</fnm></au>
    <au><snm>{Deane}</snm><fnm>A. E.</fnm></au>
  </aug>
  <source>J.\ Comput.\ Phys.</source>
  <pubdate>2009</pubdate>
  <volume>228</volume>
  <fpage>952</fpage>
  <lpage>975</lpage>
</bibl>

<bibl id="B38">
  <title><p>{A solution accurate, efficient and stable unsplit staggered mesh
  scheme for three dimensional magnetohydrodynamics}</p></title>
  <aug>
    <au><snm>{Lee}</snm><fnm>D.</fnm></au>
  </aug>
  <source>J. Comp. Phys.</source>
  <pubdate>2013</pubdate>
  <volume>243</volume>
  <fpage>269</fpage>
</bibl>

<bibl id="B39">
  <title><p>{An improved weighted essentially non-oscillatory scheme for
  hyperbolic conservation laws}</p></title>
  <aug>
    <au><snm>{Borges}</snm><fnm>R.</fnm></au>
    <au><snm>{Carmona}</snm><fnm>M.</fnm></au>
    <au><snm>{Costa}</snm><fnm>B.</fnm></au>
    <au><snm>{Don}</snm><fnm>W. S.</fnm></au>
  </aug>
  <source>Journal of Computational Physics</source>
  <pubdate>2008</pubdate>
  <volume>227</volume>
  <fpage>3191</fpage>
  <lpage>3211</lpage>
</bibl>

<bibl id="B40">
  <title><p>{On Godunov-Type Methods for Gas Dynamics}</p></title>
  <aug>
    <au><snm>{Einfeldt}</snm><fnm>B.</fnm></au>
  </aug>
  <source>SIAM Journal on Numerical Analysis</source>
  <pubdate>1988</pubdate>
  <volume>25</volume>
  <fpage>294</fpage>
  <lpage>318</lpage>
</bibl>

<bibl id="B41">
  <title><p>{Restoration of the contact surface in the HLL-Riemann
  solver}</p></title>
  <aug>
    <au><snm>{Toro}</snm><fnm>E. F.</fnm></au>
    <au><snm>{Spruce}</snm><fnm>M.</fnm></au>
    <au><snm>{Speares}</snm><fnm>W.</fnm></au>
  </aug>
  <source>Shock Waves</source>
  <pubdate>1994</pubdate>
  <volume>4</volume>
  <fpage>25</fpage>
  <lpage>34</lpage>
</bibl>

<bibl id="B42">
  <title><p>On the Theory of the Brownian Motion</p></title>
  <aug>
    <au><snm>Uhlenbeck</snm><fnm>G.</fnm></au>
    <au><snm>Ornstein</snm><fnm>L.</fnm></au>
  </aug>
  <source>Phys. Rev.</source>
  <publisher>American Physical Society</publisher>
  <pubdate>1930</pubdate>
  <volume>36</volume>
  <fpage>823</fpage>
  <lpage>-841</lpage>
  <url>http://link.aps.org/doi/10.1103/PhysRev.36.823</url>
</bibl>

<bibl id="B43">
  <title><p>{An examination of forcing in direct numerical simulations of
  turbulence}</p></title>
  <aug>
    <au><snm>Eswaran</snm><fnm>V.</fnm></au>
    <au><snm>Pope</snm><fnm>SB</fnm></au>
  </aug>
  <source>Computers \& Fluids</source>
  <publisher>Elsevier</publisher>
  <pubdate>1988</pubdate>
  <volume>16</volume>
  <issue>3</issue>
  <fpage>257</fpage>
  <lpage>-278</lpage>
  <url>http://www.sciencedirect.com/science/article/pii/0045793088900138</url>
</bibl>

<bibl id="B44">
  <title><p>{Comparing the statistics of interstellar turbulence in simulations
  and observations}</p></title>
  <aug>
    <au><snm>Federrath</snm><fnm>C</fnm></au>
    <au><snm>Roman Duval</snm><fnm>J.</fnm></au>
    <au><snm>Klessen</snm><fnm>R S</fnm></au>
    <au><snm>Schmidt</snm><fnm>W</fnm></au>
    <au><snm>{Mac Low}</snm><fnm>M. M.</fnm></au>
  </aug>
  <source>Astronomy and Astrophysics</source>
  <pubdate>2010</pubdate>
  <volume>512</volume>
  <fpage>A81</fpage>
  <url>http://www.aanda.org/10.1051/0004-6361/200912437</url>
</bibl>

<bibl id="B45">
  <title><p>{Anelastic versus Fully Compressible Turbulent Rayleigh-B{\'e}nard
  Convection}</p></title>
  <aug>
    <au><snm>{Verhoeven}</snm><fnm>J.</fnm></au>
    <au><snm>{Wieseh{\"o}fer}</snm><fnm>T.</fnm></au>
    <au><snm>{Stellmach}</snm><fnm>S.</fnm></au>
  </aug>
  <source>\apj</source>
  <pubdate>2015</pubdate>
  <volume>805</volume>
  <fpage>62</fpage>
</bibl>

<bibl id="B46">
  <title><p>{A Global Turbulence Model for Neutrino-driven Convection in
  Core-collapse Supernovae}</p></title>
  <aug>
    <au><snm>{Murphy}</snm><fnm>J. W.</fnm></au>
    <au><snm>{Meakin}</snm><fnm>C.</fnm></au>
  </aug>
  <source>\apj</source>
  <pubdate>2011</pubdate>
  <volume>742</volume>
  <fpage>74</fpage>
</bibl>

<bibl id="B47">
  <title><p>{Toward Connecting Core-collapse Supernova Theory with
  Observations. I. Shock Revival in a 15 M$_{\odot}$ Blue Supergiant Progenitor
  with SN 1987A Energetics}</p></title>
  <aug>
    <au><snm>{Handy}</snm><fnm>T.</fnm></au>
    <au><snm>{Plewa}</snm><fnm>T.</fnm></au>
    <au><snm>{Odrzywo{\l}ek}</snm><fnm>A.</fnm></au>
  </aug>
  <source>\apj</source>
  <pubdate>2014</pubdate>
  <volume>783</volume>
  <fpage>125</fpage>
</bibl>

<bibl id="B48">
  <title><p>{Revival of The Stalled Core-Collapse Supernova Shock Triggered by
  Precollapse Asphericity in the Progenitor Star}</p></title>
  <aug>
    <au><snm>{Couch}</snm><fnm>S. M.</fnm></au>
    <au><snm>{Ott}</snm><fnm>C. D.</fnm></au>
  </aug>
  <source>\apjl</source>
  <pubdate>2013</pubdate>
  <volume>778</volume>
  <fpage>L7</fpage>
</bibl>

<bibl id="B49">
  <title><p>{Non-radial instabilities and progenitor asphericities in
  core-collapse supernovae}</p></title>
  <aug>
    <au><snm>{M{\"u}ller}</snm><fnm>B.</fnm></au>
    <au><snm>{Janka}</snm><fnm>H. T.</fnm></au>
  </aug>
  <source>\mnras</source>
  <pubdate>2015</pubdate>
  <volume>448</volume>
  <fpage>2141</fpage>
  <lpage>2174</lpage>
</bibl>

<bibl id="B50">
  <title><p>Turbulence: The Legacy of A.~N.~Kolmogorov</p></title>
  <aug>
    <au><snm>Frisch</snm><fnm>U</fnm></au>
  </aug>
  <publisher>Cambridge University Press</publisher>
  <pubdate>1996</pubdate>
</bibl>

<bibl id="B51">
  <title><p>{Monotonically Integrated Large Eddy Simulation of Free Shear
  Flows}</p></title>
  <aug>
    <au><snm>Fureby</snm><fnm>C</fnm></au>
    <au><snm>Grinstein</snm><fnm>F F</fnm></au>
  </aug>
  <source>AIAA Journal</source>
  <publisher>American Institute of Aeronautics and Astronautics</publisher>
  <pubdate>1999</pubdate>
  <volume>37</volume>
  <issue>5</issue>
  <fpage>544</fpage>
  <lpage>-556</lpage>
  <url>http://dx.doi.org/10.2514/2.772</url>
</bibl>

<bibl id="B52">
  <title><p>{Analysis of implicit LES methods}</p></title>
  <aug>
    <au><snm>Aspden</snm><fnm>A</fnm></au>
    <au><snm>Nikiforakis</snm><fnm>N</fnm></au>
    <au><snm>Dalziel</snm><fnm>S</fnm></au>
    <au><snm>Bell</snm><fnm>J</fnm></au>
  </aug>
  <source>CAMCS</source>
  <pubdate>2009</pubdate>
  <volume>3</volume>
  <issue>1</issue>
  <fpage>103</fpage>
  <url>http://msp.org/camcos/2008/3-1/p06.xhtml</url>
</bibl>

<bibl id="B53">
  <title><p>{Estimating the effective Reynolds number in implicit large-eddy
  simulation}</p></title>
  <aug>
    <au><snm>Zhou</snm><fnm>Y</fnm></au>
    <au><snm>Grinstein</snm><fnm>FF</fnm></au>
    <au><snm>Wachtor</snm><fnm>AJ</fnm></au>
    <au><snm>Haines</snm><fnm>BM</fnm></au>
  </aug>
  <source>\pre</source>
  <pubdate>2014</pubdate>
  <volume>89</volume>
  <issue>1</issue>
  <fpage>013303</fpage>
  <url>http://link.aps.org/doi/10.1103/PhysRevE.89.013303</url>
</bibl>

<bibl id="B54">
  <title><p>{Anisotropy in turbulent flows and in turbulent
  transport}</p></title>
  <aug>
    <au><snm>Biferale</snm><fnm>L</fnm></au>
    <au><snm>Procaccia</snm><fnm>I</fnm></au>
  </aug>
  <source>Physics Reports</source>
  <pubdate>2005</pubdate>
  <volume>414</volume>
  <issue>2-3</issue>
  <fpage>43</fpage>
  <lpage>-164</lpage>
  <url>http://linkinghub.elsevier.com/retrieve/pii/S0370157305001547</url>
</bibl>

<bibl id="B55">
  <title><p>{Dimensional Dependence of the Hydrodynamics of Core-collapse
  Supernovae}</p></title>
  <aug>
    <au><snm>{Dolence}</snm><fnm>J. C.</fnm></au>
    <au><snm>{Burrows}</snm><fnm>A.</fnm></au>
    <au><snm>{Murphy}</snm><fnm>J. W.</fnm></au>
    <au><snm>{Nordhaus}</snm><fnm>J.</fnm></au>
  </aug>
  <source>\apj</source>
  <pubdate>2013</pubdate>
  <volume>765</volume>
  <fpage>110</fpage>
</bibl>

<bibl id="B56">
  <title><p>{High-resolution simulations of compressible convection using the
  piecewise-parabolic method}</p></title>
  <aug>
    <au><snm>{Porter}</snm><fnm>D. H.</fnm></au>
    <au><snm>{Woodward}</snm><fnm>P. R.</fnm></au>
  </aug>
  <source>\apjs</source>
  <pubdate>1994</pubdate>
  <volume>93</volume>
  <fpage>309</fpage>
</bibl>

<bibl id="B57">
  <title><p>{The satial structure and statistical properties of homogeneous
  turbulence}</p></title>
  <aug>
    <au><snm>Vincent</snm><fnm>A.</fnm></au>
    <au><snm>Meneguzzi</snm><fnm>M.</fnm></au>
  </aug>
  <source>Journal of Fluid Mechanics</source>
  <pubdate>1991</pubdate>
  <volume>225</volume>
  <fpage>1</fpage>
  <url>http://www.journals.cambridge.org/abstract\_S0022112091001957</url>
</bibl>

<bibl id="B58">
  <title><p>{Small-scale statistics in high-resolution direct numerical
  simulation of turbulence: Reynolds number dependence of one-point velocity
  gradient statistics}</p></title>
  <aug>
    <au><snm>Ishihara</snm><fnm>T.</fnm></au>
    <au><snm>Kaneda</snm><fnm>Y.</fnm></au>
    <au><snm>Yokokawa</snm><fnm>M.</fnm></au>
    <au><snm>Itakura</snm><fnm>K.</fnm></au>
    <au><snm>Uno</snm><fnm>A.</fnm></au>
  </aug>
  <source>Journal of Fluid Mechanics</source>
  <pubdate>2007</pubdate>
  <volume>592</volume>
  <fpage>335</fpage>
  <lpage>-366</lpage>
  <url>http://www.journals.cambridge.org/abstract\_S0022112007008531</url>
</bibl>

<bibl id="B59">
  <title><p>{Energy dissipation rate and energy spectrum in high resolution
  direct numerical simulations of turbulence in a periodic box}</p></title>
  <aug>
    <au><snm>{Kaneda}</snm><fnm>Y.</fnm></au>
    <au><snm>{Ishihara}</snm><fnm>T.</fnm></au>
    <au><snm>{Yokokawa}</snm><fnm>M.</fnm></au>
    <au><snm>{Itakura}</snm><fnm>K.</fnm></au>
    <au><snm>{Uno}</snm><fnm>A.</fnm></au>
  </aug>
  <source>Phys. Fluids</source>
  <pubdate>2003</pubdate>
  <volume>15</volume>
  <fpage>L21</fpage>
</bibl>

<bibl id="B60">
  <title><p>Velocity field statistics in homogeneous steady turbulence obtained
  using a high-resolution direct numerical simulation</p></title>
  <aug>
    <au><snm>Gotoh</snm><fnm>T</fnm></au>
    <au><snm>Fukayama</snm><fnm>D</fnm></au>
    <au><snm>Nakano</snm><fnm>T</fnm></au>
  </aug>
  <source>Physics of Fluids</source>
  <pubdate>2002</pubdate>
  <volume>14</volume>
  <issue>3</issue>
  <fpage>1065</fpage>
</bibl>

<bibl id="B61">
  <title><p>{On the universality of supersonic turbulence}</p></title>
  <aug>
    <au><snm>Federrath</snm><fnm>C.</fnm></au>
  </aug>
  <source>\mnras</source>
  <pubdate>2013</pubdate>
  <volume>436</volume>
  <issue>2</issue>
  <fpage>1245</fpage>
  <lpage>-1257</lpage>
  <url>http://mnras.oxfordjournals.org/cgi/doi/10.1093/mnras/stt1644</url>
</bibl>

<bibl id="B62">
  <title><p>{Measures of intermittency in driven supersonic flows}</p></title>
  <aug>
    <au><snm>Porter</snm><fnm>D.</fnm></au>
    <au><snm>Pouquet</snm><fnm>A.</fnm></au>
    <au><snm>Woodward</snm><fnm>P.</fnm></au>
  </aug>
  <source>Physical Review E</source>
  <pubdate>2002</pubdate>
  <volume>66</volume>
  <issue>2</issue>
  <fpage>1</fpage>
  <lpage>-12</lpage>
  <url>http://link.aps.org/doi/10.1103/PhysRevE.66.026301</url>
</bibl>

<bibl id="B63">
  <title><p>{Intermittency and Universality in Fully Developed Inviscid and
  Weakly Compressible Turbulent Flows}</p></title>
  <aug>
    <au><snm>{Benzi}</snm><fnm>R.</fnm></au>
    <au><snm>{Biferale}</snm><fnm>L.</fnm></au>
    <au><snm>{Fisher}</snm><fnm>R. T.</fnm></au>
    <au><snm>{Kadanoff}</snm><fnm>L. P.</fnm></au>
    <au><snm>{Lamb}</snm><fnm>D. Q.</fnm></au>
    <au><snm>{Toschi}</snm><fnm>F.</fnm></au>
  </aug>
  <source>\prl</source>
  <pubdate>2008</pubdate>
  <volume>100</volume>
  <issue>23</issue>
  <fpage>234503</fpage>
</bibl>

<bibl id="B64">
  <title><p>{Anomalous and dimensional scaling in anisotropic
  turbulence}</p></title>
  <aug>
    <au><snm>Biferale</snm><fnm>L.</fnm></au>
    <au><snm>Daumont</snm><fnm>I.</fnm></au>
    <au><snm>Lanotte</snm><fnm>a</fnm></au>
    <au><snm>Toschi</snm><fnm>F.</fnm></au>
  </aug>
  <source>Physical Review E</source>
  <pubdate>2002</pubdate>
  <volume>66</volume>
  <issue>5</issue>
  <fpage>056306</fpage>
  <url>http://link.aps.org/doi/10.1103/PhysRevE.66.056306</url>
</bibl>

<bibl id="B65">
  <title><p>Evidences of Bolgiano-Obhukhov scaling in three-dimensional
  Rayleigh-B\'enard convection</p></title>
  <aug>
    <au><snm>Calzavarini</snm><fnm>E</fnm></au>
    <au><snm>Toschi</snm><fnm>F</fnm></au>
    <au><snm>Tripiccione</snm><fnm>R</fnm></au>
  </aug>
  <source>Phys. Rev. E</source>
  <publisher>American Physical Society</publisher>
  <pubdate>2002</pubdate>
  <volume>66</volume>
  <fpage>016304</fpage>
  <url>http://link.aps.org/doi/10.1103/PhysRevE.66.016304</url>
</bibl>

<bibl id="B66">
  <title><p>{Universality of anisotropic fluctuations from numerical
  simulations of turbulent flows}</p></title>
  <aug>
    <au><snm>Biferale</snm><fnm>L</fnm></au>
    <au><snm>Calzavarini</snm><fnm>E</fnm></au>
    <au><snm>Toschi</snm><fnm>F</fnm></au>
    <au><snm>Tripiccione</snm><fnm>R</fnm></au>
  </aug>
  <source>Europhysics Letters (EPL)</source>
  <pubdate>2003</pubdate>
  <volume>64</volume>
  <issue>4</issue>
  <fpage>461</fpage>
  <lpage>-467</lpage>
  <url>http://stacks.iop.org/0295-5075/64/i=4/a=461?key=crossref.8d22e8a90fa965d7b3014481e1962416</url>
</bibl>

<bibl id="B67">
  <title><p>{Examination of Kolmogorov's 4/5 Law by High-Resolution Direct
  Numerical Simulation Data of Turbulence}</p></title>
  <aug>
    <au><snm>Kaneda</snm><fnm>Y</fnm></au>
    <au><snm>Yoshino</snm><fnm>J</fnm></au>
    <au><snm>Ishihara</snm><fnm>T</fnm></au>
  </aug>
  <source>Journal of the Physical Society of Japan</source>
  <pubdate>2008</pubdate>
  <volume>77</volume>
  <issue>6</issue>
  <fpage>064401</fpage>
  <url>http://journals.jps.jp/doi/abs/10.1143/JPSJ.77.064401</url>
</bibl>

<bibl id="B68">
  <title><p>{Small-Scale Properties of Turbulent Rayleigh-B\'{e}nard
  Convection}</p></title>
  <aug>
    <au><snm>Lohse</snm><fnm>D</fnm></au>
    <au><snm>Xia</snm><fnm>KQ</fnm></au>
  </aug>
  <source>Annual Review of Fluid Mechanics</source>
  <pubdate>2010</pubdate>
  <volume>42</volume>
  <issue>1</issue>
  <fpage>335</fpage>
  <lpage>-364</lpage>
  <url>http://www.annualreviews.org/doi/abs/10.1146/annurev.fluid.010908.165152</url>
</bibl>

<bibl id="B69">
  <title><p>{Dissipation and enstrophy in isotropic turbulence: Resolution
  effects and scaling in direct numerical simulations}</p></title>
  <aug>
    <au><snm>Donzis</snm><fnm>Da</fnm></au>
    <au><snm>Yeung</snm><fnm>P. K.</fnm></au>
    <au><snm>Sreenivasan</snm><fnm>K. R.</fnm></au>
  </aug>
  <source>Phys. Fluids</source>
  <pubdate>2008</pubdate>
  <volume>20</volume>
  <issue>4</issue>
  <fpage>045108</fpage>
  <url>http://scitation.aip.org/content/aip/journal/pof2/20/4/10.1063/1.2907227</url>
</bibl>

<bibl id="B70">
  <title><p>{General-relativistic Simulations of Three-dimensional
  Core-collapse Supernovae}</p></title>
  <aug>
    <au><snm>{Ott}</snm><fnm>C. D.</fnm></au>
    <au><snm>{Abdikamalov}</snm><fnm>E.</fnm></au>
    <au><snm>{M{\"o}sta}</snm><fnm>P.</fnm></au>
    <au><snm>{Haas}</snm><fnm>R.</fnm></au>
    <au><snm>{Drasco}</snm><fnm>S.</fnm></au>
    <au><snm>{O'Connor}</snm><fnm>E. P.</fnm></au>
    <au><snm>{Reisswig}</snm><fnm>C.</fnm></au>
    <au><snm>{Meakin}</snm><fnm>C. A.</fnm></au>
    <au><snm>{Schnetter}</snm><fnm>E.</fnm></au>
  </aug>
  <source>\apj</source>
  <pubdate>2013</pubdate>
  <volume>768</volume>
  <fpage>115</fpage>
</bibl>

<bibl id="B71">
  <title><p>{An adaptive local deconvolution method for implicit
  LES}</p></title>
  <aug>
    <au><snm>Hickel</snm><fnm>S</fnm></au>
    <au><snm>Adams</snm><fnm>Na</fnm></au>
    <au><snm>Domaradzki</snm><fnm>JA</fnm></au>
  </aug>
  <source>Journal of Computational Physics</source>
  <pubdate>2006</pubdate>
  <volume>213</volume>
  <issue>1</issue>
  <fpage>413</fpage>
  <lpage>-436</lpage>
  <url>http://linkinghub.elsevier.com/retrieve/pii/S0021999105003864</url>
</bibl>

<bibl id="B72">
  <title><p>{A bandwidth-optimized WENO scheme for the effective direct
  numerical simulation of compressible turbulence}</p></title>
  <aug>
    <au><snm>Mart\'{\i}n</snm><fnm>M.P.</fnm></au>
    <au><snm>Taylor</snm><fnm>E.M.</fnm></au>
    <au><snm>Wu</snm><fnm>M.</fnm></au>
    <au><snm>Weirs</snm><fnm>V.G.</fnm></au>
  </aug>
  <source>Journal of Computational Physics</source>
  <pubdate>2006</pubdate>
  <volume>220</volume>
  <issue>1</issue>
  <fpage>270</fpage>
  <lpage>-289</lpage>
  <url>http://linkinghub.elsevier.com/retrieve/pii/S0021999106002312</url>
</bibl>

<bibl id="B73">
  <title><p>{An improved reconstruction method for compressible flows with low
  Mach number features}</p></title>
  <aug>
    <au><snm>Thornber</snm><fnm>B.</fnm></au>
    <au><snm>Mosedale</snm><fnm>a</fnm></au>
    <au><snm>Drikakis</snm><fnm>D.</fnm></au>
    <au><snm>Youngs</snm><fnm>D.</fnm></au>
    <au><snm>Williams</snm><fnm>R.J.R.</fnm></au>
  </aug>
  <source>Journal of Computational Physics</source>
  <pubdate>2008</pubdate>
  <volume>227</volume>
  <issue>10</issue>
  <fpage>4873</fpage>
  <lpage>-4894</lpage>
  <url>http://linkinghub.elsevier.com/retrieve/pii/S0021999108000429</url>
</bibl>

<bibl id="B74">
  <title><p>{Supernova mechanisms}</p></title>
  <aug>
    <au><snm>{Bethe}</snm><fnm>H. A.</fnm></au>
  </aug>
  <source>Rev. Mod. Phys.</source>
  <pubdate>1990</pubdate>
  <volume>62</volume>
  <fpage>801</fpage>
</bibl>

<bibl id="B75">
  <title><p>{Black Hole Formation in Failing Core-Collapse
  Supernovae}</p></title>
  <aug>
    <au><snm>{O'Connor}</snm><fnm>E.</fnm></au>
    <au><snm>{Ott}</snm><fnm>C. D.</fnm></au>
  </aug>
  <source>\apj</source>
  <pubdate>2011</pubdate>
  <volume>730</volume>
  <fpage>70</fpage>
</bibl>

<bibl id="B76">
  <title><p>{The Physics of the Neutrino Mechanism of Core-collapse
  Supernovae}</p></title>
  <aug>
    <au><snm>{Pejcha}</snm><fnm>O.</fnm></au>
    <au><snm>{Thompson}</snm><fnm>T. A.</fnm></au>
  </aug>
  <source>\apj</source>
  <pubdate>2012</pubdate>
  <volume>746</volume>
  <fpage>106</fpage>
</bibl>

</refgrp>
} 
\end{backmatter}

\acrodef{CCSN}{core-collapse supernovae}
\acrodef{DNS}{direct numerical simulations}
\acrodef{HRSC}{high resolution shock capturing}
\acrodef{ILES}{implicit large eddy simulation}
\acrodef{MILES}{monotone integrated large eddy simulation}
\acrodef{MC}{monotonized central}
\acrodef{MUSCL}{monotonic upstream-centered scheme for conservation laws}
\acrodef{PPM}{piecewise parabolic method}
\acrodef{PSD}{power spectral density}
\acrodef{PNS}{proto-neutron star}
\acrodef{RMS}{root mean square}

\end{document}